\documentclass[camera,letterpaper,nomarginnotes,nonarrowgutter]{jpaper} 

\usepackage[sort,compress]{cite}
\usepackage{amsmath,amssymb,amsfonts}
\usepackage{algorithmic}
\usepackage{graphicx}
\usepackage{textcomp}
\usepackage[bookmarks=true,breaklinks=true,letterpaper=true,colorlinks,linkcolor=black,citecolor=black,urlcolor=blue]{hyperref}
\usepackage{balance} 
\usepackage{fancyhdr}
\usepackage{datetime}
\usepackage{enumitem}
\usepackage{marginnote}
\usepackage{soul}
\usepackage{dblfloatfix}    

\usepackage{xspace}
\usepackage{xcolor}
\usepackage{pifont}
\usepackage{listings}
\usepackage{siunitx}
\usepackage{setspace}
\usepackage{multirow}
\usepackage{caption}

\makeatletter
\g@addto@macro{\normalsize}{%
  \setlength{\abovedisplayskip}{3pt plus 0.5pt minus 1pt}
  \setlength{\belowdisplayskip}{3pt plus 0.5pt minus 1pt}
  \setlength{\abovedisplayshortskip}{0pt}
  \setlength{\belowdisplayshortskip}{0pt}
  \setlength{\intextsep}{4pt plus 1pt minus 1pt}
  \setlength{\textfloatsep}{4pt plus 1pt minus 1pt}
  \setlength{\skip\footins}{5pt plus 1pt minus 1pt}}
\makeatother

\usepackage{titlesec}
\titlespacing\section{0pt}{2pt plus 1pt minus 1pt}{3pt plus 1pt minus 2pt}
\titlespacing\subsection{0pt}{2pt plus 1pt minus 1pt}{3pt plus 1pt minus 2pt}
\titlespacing\subsubsection{0pt}{2pt plus 1pt minus 1pt}{3pt plus 1pt minus 2pt}

\newcommand{\affilETH}[0]{\textsuperscript{\S}}
\newcommand{\affilETU}[0]{\textsuperscript{$\dagger$}}
\newcommand{\affilGSC}[0]{\textsuperscript{$\odot$}}

\newif\ifonur
\newif\ifsubmission
\newif\ifrevision
\newif\ifdefault
\newif\iftcadrev
\newif\ifcameraready
\newif\ifversioning
\newif\ifextendedversionmark
\onurtrue
\submissionfalse
\revisionfalse
\defaultfalse
\tcadrevfalse
\camerareadyfalse
\versioningfalse
\extendedversionmarktrue
\newcommand\X[0]{{DRAM Bender\xspace}}

\newcommand{\One}{\emph{(i)}~}
\newcommand{\Two}{\emph{(ii)}~}
\newcommand{\Three}{\emph{(iii)}~}
\newcommand{\Four}{\emph{(iv)}~}
\newcommand{\Five}{\emph{(v)}~}

\newcommand{\cmark}{\ding{51}}
\newcommand{\xmark}{\ding{55}}

\newcommand{\squishlist}{
 \begin{list}{$\circ$}
  { \setlength{\itemsep}{0pt}
     \setlength{\parsep}{0pt}
     \setlength{\topsep}{0pt}
     \setlength{\partopsep}{0pt}
     \setlength{\leftmargin}{1em}
     \setlength{\labelwidth}{1em}
     \setlength{\labelsep}{0.5em} } }

\newcommand{\squishsublist}{
\begin{list}{$\rightarrow$}
 { \setlength{\itemsep}{0pt}
    \setlength{\parsep}{0pt}
    \setlength{\topsep}{-10em}
    \setlength{\partopsep}{-3pt}
    \setlength{\leftmargin}{1em}
    \setlength{\labelwidth}{1em}
    \setlength{\labelsep}{0.5em} } }

\newcommand{\squishend}{
  \end{list}  }

\ifsubmission
    \newcommand\del[1]{\textcolor{red}{#1}}
    \newcommand\new[1]{\textcolor{blue}{#1}}
    
    \newcommand\agy[1]{{#1}}
    \newcommand\yct[1]{{#1}}
    \newcommand\lois[1]{{#1}}
    \newcommand\hh[1]{{#1}}
    \newcommand\atb[1]{{#1}}
    \newcommand\om[1]{{#1}}
    \newcommand{\hluo}[1]{{#1}}
    
    \newcommand\atbc[1]{}
    \newcommand\loiscomment[1]{}
    \newcommand\agycomment[1]{}
    \newcommand{\hluocomment}[1]{}
\fi
\ifrevision

    \definecolor{gold}{rgb}{0.83, 0.69, 0.22}
    \definecolor{brightpink}{rgb}{1.0, 0.0, 0.5}
    \definecolor{ao}{rgb}{0.0, 0.5, 0.0}

    \newcommand\agy[1]{\textcolor{black}{#1}}
    \newcommand\yct[1]{\textcolor{black}{#1}}
    \newcommand\lois[1]{\textcolor{black}{#1}}
    \newcommand\hh[1]{\textcolor{black}{#1}}
    \newcommand\atb[1]{\textcolor{black}{#1}}
    \newcommand{\hluo}[1]{\textcolor{black}{#1}}
    
    \newcommand\atbc[1]{}
    \newcommand\loiscomment[1]{}
    \newcommand\agycomment[1]{}
    \newcommand{\hluocomment}[1]{}
\fi
\ifonur
    \newcommand\del[1]{}
    \newcommand\new[1]{{#1}}
    \newcommand\om[1]{{#1}}
    \newcommand\agy[1]{{#1}}
    \newcommand\yct[1]{{#1}}
    \newcommand\lois[1]{{#1}}
    \newcommand\hh[1]{{#1}}
    \newcommand\atb[1]{{#1}}
    \newcommand{\hluo}[1]{{#1}}
    
    \newcommand\atbc[1]{}
    \newcommand\loiscomment[1]{}
    \newcommand\agycomment[1]{}
    \newcommand{\hluocomment}[1]{}
\fi
\ifdefault

        \newcommand\om[1]{\textcolor{black}{#1}}

    \newcommand\agycomment[1]{\textcolor{magenta}{\textbf{@gy:} #1}}
    \definecolor{amber}{rgb}{1.0, 0.75, 0.0}
    \newcommand\agy[1]{\textcolor{amber}{#1}}
    \newcommand\yct[1]{\textcolor{olive}{#1}}
    \newcommand\lois[1]{\textcolor{blue}{#1}}
    \newcommand\loiscomment[1]{\textcolor{red}{\textbf{@lois:} #1}}
    \newcommand\hh[1]{\textcolor{cyan}{#1}}
    \newcommand\atbc[1]{\textcolor{purple}{\textbf{@atb:} #1}}
    \newcommand\atb[1]{\textcolor{purple}{#1}}
    
    \definecolor{moegi}{rgb}{0.357, 0.537, 0.188}
    \newcommand{\hluo}[1]{\textcolor{moegi}{#1}}
    \newcommand{\hluocomment}[1]{\textcolor{moegi}{\textit{[@hluo: #1]}}}
\fi

\usepackage{xargs}                    
\usepackage{titlesec}
\usepackage{titletoc}
\usepackage{clipboard}
\usepackage[all]{nowidow}
\usepackage[framemethod=tikz]{mdframed}
\usepackage[most]{tcolorbox}
\tcbuselibrary{breakable}
\tcbuselibrary{hooks}

\ifcameraready

\newcommandx{\changev}[2][1=]{}

\newcommand{\omcr}[1]{\textcolor{purple}{#1}}
\usepackage[colorinlistoftodos,prependcaption,textsize=small]{todonotes}
\paperwidth=\dimexpr \paperwidth + 4cm\relax
\oddsidemargin=\dimexpr\oddsidemargin + 2cm\relax
\evensidemargin=\dimexpr\evensidemargin + 2cm\relax
\marginparwidth=\dimexpr \marginparwidth + 2cm\relax
\newcommand{\atbcommentside}[1]{\todo[size=\scriptsize, linecolor=orange, bordercolor=orange, backgroundcolor=white]{\textcolor{blue}{\textbf{@atb:} #1}}}
\else
\newcommandx{\changev}[2][1=]{}

\newcommand{\omcr}[1]{{#1}}
\newcommand{\atbcommentside}[1]{}
\usepackage[colorinlistoftodos,prependcaption,textsize=small]{todonotes}
\fi

\iftcadrev
\newcommand{\tcadreva}[1]{\textcolor{red}{#1}}
\definecolor{dark-green}{rgb}{0.00, 0.45, 0.00}
\newcommand{\tcadrevb}[1]{\textcolor{dark-green}{#1}}
\definecolor{goldbutdark}{rgb}{0.85, 0.65, 0.12}
\newcommand{\tcadrevc}[1]{\textcolor{goldbutdark}{#1}}
\newcommand{\tcadrevcommon}[1]{\textcolor{blue}{#1}}
\newcommand{\revdel}[1]{}
\else
\newcommand{\tcadreva}[1]{#1}
\definecolor{dark-green}{rgb}{0.00, 0.45, 0.00}
\newcommand{\tcadrevb}[1]{#1}
\definecolor{goldbutdark}{rgb}{0.85, 0.65, 0.12}
\newcommand{\tcadrevc}[1]{#1}
\newcommand{\tcadrevcommon}[1]{#1}
\newcommand{\revdel}[1]{}
\fi

\ifextendedversionmark
\newcommand{\ext}[1]{\textcolor{blue}{#1}}
\else
\newcommand{\ext}[1]{#1}
\fi

\newcommand{\exploitingRowHammerAllCitations}[0]{\cite{fournaris2017exploiting, poddebniak2018attacking, tatar2018throwhammer, carre2018openssl, barenghi2018software, zhang2018triggering, bhattacharya2018advanced, google-project-zero, kim2014flipping, rowhammergithub, seaborn2015exploiting, van2016drammer, gruss2016rowhammer, razavi2016flip, pessl2016drama, xiao2016one, bosman2016dedup, bhattacharya2016curious, burleson2016invited, qiao2016new, brasser2017can, jang2017sgx, aga2017good, mutlu2017rowhammer, tatar2018defeating, gruss2018another, lipp2018nethammer, van2018guardion, frigo2018grand, cojocar2019eccploit,  ji2019pinpoint, mutlu2019rowhammer, hong2019terminal, kwong2020rambleed, frigo2020trrespass, cojocar2020rowhammer, weissman2020jackhammer, zhang2020pthammer, yao2020deephammer, deridder2021smash, hassan2021utrr, jattke2022blacksmith, tol2022toward, kogler2022half, orosa2022spyhammer, zhang2022implicit, liu2022generating, cohen2022hammerscope, zheng2022trojvit, fahr2022frodo, tobah2022spechammer, rakin2022deepsteal, aydin2022cyber, mus2022jolt, wang2022research, lefforge2023reverse,fahr2022effects, kaur2022work, cai2022feasibility, li2022cyberradar, roohi2022efficient, staudigl2022neurohammer, yang2022socially, islam2022signature}}

\newcommand{\mitigatingRowHammerAllCitations}[0]{\cite{AppleRefInc, rh-hp,rh-lenovo,greenfield2012throttling, kim2014flipping, kim2014architectural, bains14c, aweke2016anvil, bains2015row, bains2016row, bains2016distributed, son2017making, seyedzadeh2018cbt,irazoqui2016mascat, you2019mrloc, lee2018twice, park2020graphene, yaglikci2021security, yaglikci2021blockhammer, frigo2020trrespass, kang2020cat, hassan2021utrr, qureshi2022hydra, saileshwar2022randomized, brasser2017can, konoth2018zebram, van2018guardion, vig2018rapid,  kim2022mithril, lee2021cryoguard, marazzi2022protrr, zhang2022softtrr, joardar2022learning, juffinger2023csi, yaglikci2022hira, saxena2022aqua, enomoto2022efficient, manzhosov2022revisiting, ajorpaz2022evax, naseredini2022alarm, joardar2022machine, hassan2022case, zhang2020leveraging,loughlin2021stop, devaux2021method, han2021surround, fakhrzadehgan2022safeguard, saroiu2022price, saroiu2022configure, loughlin2022moesiprime, zhou2022lt, hong2023dsac, mutlu2023fundamentally, marazzi2023rega, di2023copy, sharma2022review, woo2023scalable, park2022row, wi2023shadow, kim2023ddr5, gude2023defending, guha2022criticality, france2022modeling, france2022reducing, bennett2021panopticon, enomoto2022efficient, arikan2022processor, tomita2022extracting}}

\newcommand{\pimcitations}[0]
{\cite{stone1970logic, Kautz1969, shaw1981non, kogge1994, gokhale1995processing, patterson1997case, oskin1998active, kang1999flexram, Mai:2000:SMM:339647.339673,murphy2001characterization, Draper:2002:ADP:514191.514197,aga.hpca17,eckert2018neural,fujiki2019duality,kang.icassp14,seshadri2017ambit,seshadri.arxiv16,seshadri2015fast,seshadri2013rowclone,angizi2019graphide,kim.hpca18,kim.hpca19,gao2019computedram,chang2016low,xin2020elp2im,li.micro17,deng.dac2018,hajinazarsimdram,rezaei2020nom,wang2020figaro,ali2019memory,li.dac16,angizi2018pima,angizi2018cmp,angizi2019dna,levy.microelec14,kvatinsky.tcasii14,shafiee2016isaac,kvatinsky.iccd11,kvatinsky.tvlsi14,gaillardon2016plim,bhattacharjee2017revamp,hamdioui2015memristor,xie2015fast,hamdioui2017myth,yu2018memristive,syncron,fernandez2020natsa,cali2020genasm,kim.bmc18,ahn.pei.isca15,ahn.tesseract.isca15,boroumand.asplos18,boroumand2019conda,singh2019napel,asghari-moghaddam.micro16,DBLP:conf/sigmod/BabarinsaI15,chi2016prime,farmahini2015nda,gao.pact15,DBLP:conf/hpca/GaoK16,gu.isca16,guo2014wondp,hashemi.isca16,cont-runahead,hsieh.isca16,kim.isca16,kim.sc17,DBLP:conf/IEEEpact/LeeSK15,liu-spaa17,morad.taco15,nai2017graphpim,pattnaik.pact16,pugsley2014ndc,zhang.hpdc14,zhu2013accelerating,DBLP:conf/isca/AkinFH15,gao2017tetris,drumond2017mondrian,dai2018graphh,zhang2018graphp,huang2020heterogeneous,zhuo2019graphq,santos2017operand,ghoseibm2019,wen2017rebooting,besta2021sisa,ferreira2021pluto,olgun2021quactrng,lloyd2015memory,elliott1999computational,zheng2016tcam,landgraf2021combining,rodrigues2016scattergather,lloyd2018dse,lloyd2017keyvalue,gokhale2015rearr,nair2015active,jacob2016compiling,sura2015data,nair2015evolution,balasubramonian2014near,xi2020memory,impica,boroumand2016pim,giannoula2022sparsep,giannoula2022sigmetrics,denzler2021casper,boroumand2021polynesia,boroumand2021icde,singh2021fpga,singh2021accelerating,herruzo2021enabling,yavits2021giraf,asgarifafnir,boroumand2021google_arxiv,boroumand2021google,amiraliphd,singh2020nero,seshadri.bookchapter17,diab2022high,diab2022hicomb,fujiki2018memory,zha2020hyper,mutlu2013memory,mutlu.superfri15,ahmed2019compiler,jain2018computing,ghiasi2022genstore,deoliveira2021IEEE,deoliveira2021,cho2020mcdram,shin2018mcdram,gu2020ipim,lavenier2020,Zois2018}}

\def\UrlBreaks{\do\/\do-\/\do.\/\do:}

\expandafter\def\expandafter\UrlBreaks\expandafter{\UrlBreaks
  \do\a\do\b\do\c\do\d\do\e\do\f\do\g\do\h\do\i\do\j
  \do\k\do\l\do\m\do\n\do\o\do\p\do\q\do\r\do\s\do\t
  \do\u\do\v\do\w\do\x\do\y\do\z\do\A\do\B\do\C\do\D
  \do\E\do\F\do\G\do\H\do\I\do\J\do\K\do\L\do\M\do\N
  \do\O\do\P\do\Q\do\R\do\S\do\T\do\U\do\V\do\W\do\X
  \do\Y\do\Z}

\title{\Large{\X: An Extensible and {Versatile} FPGA-based Infrastructure\\to Easily Test State-of-the-art DRAM Chips}}

\author{
{Ataberk Olgun\affilETH}\qquad%
{Hasan Hassan\affilETH}\qquad
{A. Giray Ya\u{g}l{\i}k\c{c}{\i}\affilETH}\qquad
{Yahya Can Tu\u{g}rul\affilETH\affilETU}\qquad \\%
{Lois Orosa\affilETH\affilGSC}\qquad%
{Haocong Luo\affilETH}\qquad%
{Minesh Patel\affilETH}\qquad
{O\u{g}uz Ergin\affilETU}\qquad%
{Onur Mutlu\affilETH}\qquad\\
{\affilETH \emph{ETH Z{\"u}rich}} \qquad \affilETU \emph{TOBB ET\"U} \qquad \affilGSC \emph{Galician Supercomputing Center} %
}

\begin{document}

\definecolor{RED}{rgb}{1, 0, 0}
\definecolor{lightblue}{rgb}{0.980, 0.956, 0.623}
\newcommand{\MA}{}
\DeclareRobustCommand{\MA}[1]{{\sethlcolor{lightblue}\hl{#1}}}

\let\oldmarginnote\marginnote
\renewcommand*{\marginfont}{\tiny}
\renewcommand{\marginnote}[2][rectangle,draw,fill=blue!40,rounded corners]{%
        \oldmarginnote{%
        \tikz \node at (0,0) [#1]{#2};}%
        }
        
\definecolor{lightyellow}{rgb}{0.980, 0.956, 0.623}

\newcommand{\boxbegin} {
	\begin{tcolorbox}[enhanced, frame hidden, colback=gray!50, breakable]
}

\newcommand{\boxend} {
	\end{tcolorbox}
}

\newcommand{\yboxbegin} {
	\begin{tcolorbox}[breakable, enhanced, frame hidden, colback=yellow!50]
}

\newcommand{\yboxend} {
	\end{tcolorbox}
}

\mdfdefinestyle{graybox}{
    splittopskip=0,%
    splitbottomskip=0,%
    frametitleaboveskip=0,
    frametitlebelowskip=0,
    skipabove=0,%
    skipbelow=0,%
    leftmargin=0,%
    rightmargin=0,%
    innertopmargin=0mm,%
    innerbottommargin=0mm,%
    roundcorner=1mm,%
    backgroundcolor=lightblue,
    hidealllines=true}
    
\mdfdefinestyle{graybox2}{
    splittopskip=0,%
    splitbottomskip=0,%
    frametitleaboveskip=0,
    frametitlebelowskip=0,
    skipabove=0,%
    skipbelow=0,%
    leftmargin=0,%
    rightmargin=0,%
    innertopmargin=2mm,%
    innerbottommargin=2mm,%
    roundcorner=2mm,%
    backgroundcolor=lightblue,
    hidealllines=true}

\newcommand{\bboxbegin}{
    \begin{mdframed}[style=graybox]
}

\newcommand{\bboxend}{
    \end{mdframed}
}

\newcommand{\yyboxbegin}{
    \begin{mdframed}[style=graybox2]
}

\newcommand{\yyboxend}{
    \end{mdframed}
}

\renewcommand\citepunct{, }
\renewcommand\citeform[1]{#1}
\renewcommand\citedash{-}
\renewcommand\citeleft{[}
\renewcommand\citeright{]}
\maketitle

\ifversioning
    \fancypagestyle{firstpage}{
        \fancyhf{}
        \renewcommand{\headrulewidth}{0pt}
        \fancyhead[C]{\textcolor{blue}{TCAD Camera Ready -- Version 6.0~~[\today{}~\currenttime{}]}}
        \fancyfoot[C]{\thepage}
    }
    \thispagestyle{firstpage}
    \pagestyle{firstpage}
\else
\thispagestyle{plain}
\pagestyle{plain}
\fi


\begin{abstract}

\om{To understand and improve DRAM performance, reliability, security, and energy efficiency, prior works} study characteristics of commodity DRAM chips. \del{These characterization studies yield novel insights, motivating new mechanisms that improve the performance, {energy consumption}, {reliability{,} and security} of DRAM \om{real DRAM chips}. }{{Unfortunately, state-of-the-art} open source infrastructures capable of conducting such studies are obsolete, poorly supported, \om{or} difficult to use, or their inflexibility limits the types of studies they can conduct.}

We propose \X{}, a new FPGA-based infrastructure that enables {experimental studies on state-of-the-art DRAM chips}. \hh{\X{} offers {three} key features \om{at the same time}.} 
\om{First,} \X{} enables directly interfacing with a DRAM \om{chip} through its low-level interface. This {allows} users {to} issue DRAM commands in arbitrary order and with {finer-grained time intervals compared to other open source infrastructures.} \om{Second}, \X{} exposes easy-to-use C++ and Python programm\om{ing} interfaces\om{, allowing users to quickly and easily develop different types of DRAM experiments.} \om{Third,} {\X{} is} easily extensible. The modular design of \X{} allows extending it to (i) support existing and emerging DRAM interfaces, and (ii) run on new {commercial} or custom FPGA boards with {little} effort.

To \hh{demonstrate} that \X{} is a {versatile} \hh{infrastructure}, we conduct {three} case studies{, two of which lead to new observations about the \om{DRAM} RowHammer \om{vulnerability}}. {In particular, we show that data patterns supported by \X{} {{uncover}}\del{can uncover} a larger set of bit-flips on a victim row {than those} commonly used by prior work.}\del{Our results show that (i) the order and frequency of DRAM accesses have a significant impact on the effectiveness of RowHammer attacks, (ii) data patterns supported by \X{} can uncover a larger set of bit-flips on a victim row compared to the data patterns commonly used by prior work, (iii) and new DDR4 \om{chips} {naturally} support processing-in-memory capability.}
We demonstrate the extensibility of \X{} by implementing it on {five different FPGAs with DDR4 and DDR3 support.}\del{ \hh{four different FPGAs with DDR4 support {in addition to one FPGA board with DDR3 support}}.}
{\X{} is {freely and openly} available at \url{https://github.com/CMU-SAFARI/DRAM-Bender}.}

\end{abstract}
\section{Introduction}
\label{sec:introduction}

DRAM~\cite{dennard1968dram} {is} \hh{the dominant} technology {used in} building \hh{the} main memory \hh{of computer systems} as {it} provides \hh{low access latency} and can be manufactured at {low} cost per \hh{bit}. 
{Challenges in} DRAM technology {scaling} render continuous improvement of DRAM \om{performance}{, energy efficiency, security,} and reliability {difficult}~\cite{mutlu2013memory,mutlu2017rowhammer}. 

\om{To improve DRAM in all aspects and overcome DRAM scaling challenges, it is critical to experimentally understand the operation and characteristics of real DRAM chips. As such, it is critical to develop experimental testing infrastructures to efficiently and easily test real state-of-the-art DRAM chips.} \om{We refer to these infrastructures as DRAM testing infrastructures.} \om{These infrastructures enable {at least} two major {possibilities}.}

\noindent
\textbf{Understanding DRAM Scaling Trends.} \hh{Technology node} scaling affects various \hh{DRAM characteristics}, such as DRAM cell capacitance and capacitive crosstalk, \hh{negatively impacting} {the} reliability, latency\om{, and security {characteristics}} \hh{of DRAM}~\cite{kang-memcon2014,mandelman2002challenges,mutlu2013memory}. \hh{Therefore, it} is important to \hh{thoroughly} analyze these effects to understand how \om{real} DRAM \om{chip} {characteristics} change with {technology} scaling. \hh{Using DRAM testing infrastructures, prior studies uncover insights that inspire new mechanisms{/methodologies} for improving DRAM access latency~\cite{kim2018solar, hassan2016chargecache, chang2016understanding, lee2017design, chang2017thesis,wang2018reducing,lee2015adaptive,choi2015multiple,chang2016low,lee2013tiered,son2013reducing}, improving DRAM energy {efficiency}~\cite{chang2017understanding,qureshi2015avatar,patel2017reaper,hassan2019crow,liu2012raidr}, {understanding {and mitigating} the Rowhammer problem~\cite{kim2014flipping,kim2020revisiting,orosa2021deeper,mutlu2018rowhammer,mutlu2017rowhammer,yaglikci2022understanding}, and understanding DRAM {data} retention failures~\cite{khan2014efficacy,hamamoto1998retention,venkatesan2006retention,khan2017detecting,qureshi2015avatar,liu2013experimental,liu2012raidr,patel2017reaper}}.} \om{Such understanding is very difficult, if not impossible, to develop without testing real chips.}

\noindent
\textbf{Uncovering {Undocumented} Functionality.} Standard DRAM interfaces \hh{define} constraints \hh{(e.g., timing parameters)} that \hh{the memory controller obeys to} guarantee correct DRAM operation. \hh{These} constraints{, by design, preclude the discovery of} \hh{various} {undocumented} operations that \hh{the DRAM technology or the chip design {{may}} inherently {be} {capable of}}. \hh{For example, recent works} {demonstrate} techniques that enable \hh{several in-DRAM} computation primitives (e.g., bulk data copy~\cite{seshadri2013rowclone,gao2019computedram,olgun2021pidram}, bitwise operations~\cite{seshadri2017ambit,gao2019computedram,seshadri2015fast}, {true random number generation}~\cite{kim2019d, olgun2021quactrng,talukder2019exploiting}, and {physical unclonable functions}~\cite{kim2018dram,sutar2018d,talukder2019exploiting,orosa2021codic}) \hh{using} off-the-shelf DRAM devices by violating \hh{the \om{DRAM timing parameters}}. \om{Such undocumented DRAM behavior is possible to discover and demonstrate \emph{only} by testing real DRAM chips using infrastructures that can violate DRAM timing parameters.}

\noindent
\changev{\ref{q:r1q1}}\textbf{\tcadreva{Limitations of Existing Infrastructure{s}.}} 
\om{DRAM timing parameters \emph{cannot} be {freely modified or violated} in \tcadrevcommon{widely available} computer systems. In such systems, the CPU sends memory requests (e.g., load/store) to the memory controller to access DRAM. The memory controller translates the memory request to appropriate low-level DRAM commands that strictly comply with the interface specification (i.e., DRAM standard) of the DRAM chip (i.e., the memory controller does \emph{not} violate DRAM timing parameters). Thus, the two major research directions cannot be explored using typical computer systems as such exploration would require modifications to existing memory controllers.}

\begin{table*}[!b]
\centering
\footnotesize
\caption{Compari\om{son {of}} \X{} \hh{to the \om{major} existing {open source} DRAM testing infrastructures}}
\label{table:tools}
\begin{tabular}{|l||c|c|c|c|c|}
\hline
           \textbf{Testing \hh{Infrastructure}} & 
           \begin{tabular}[c]{@{}l@{}} \textbf{{Interface (IF) Restrictions}} \end{tabular} &
           \begin{tabular}[c]{@{}l@{}}\textbf{\hh{Ease of} Use}\end{tabular} & \begin{tabular}[c]{@{}l@{}}\textbf{\hh{Extensibility}}\end{tabular} &
           \begin{tabular}[c]{@{}l@{}}\textbf{\hh{{Protocol} Support}}\end{tabular} & \begin{tabular}[c]{@{}l@{}}\textbf{FPGA \hh{Support}}\end{tabular} \\ \hline \hline
{SoftMC~\cite{hassan2017softmc}} & Data IF & \xmark & \xmark & DDR3 & One Prototype \\ \hline
{\yct{LiteX} \lois{RowHammer} Tester (\yct{LRT})~\cite{litex.github}} & Command \& Data IF & \xmark & \cmark & DDR3/4, LPDDR4 & Two Prototypes \\ \hline
\hline
\textbf{\X{} (this work)} & \textbf{No Restrictions} & \cmark & \cmark & \textbf{DDR3/DDR4} & \textbf{Five Prototypes} \\ \hline
\end{tabular}
\end{table*}

\del{\hh{In a typical computer system, the CPU \om{sends memory requests (e.g., load/store) to the memory controller to access DRAM.} The memory controller is responsible for translating the \om{memory requests} to appropriate low-level DRAM commands that \om{strictly} comply with the interface specification \om{(i.e., DRAM standard)} of the DRAM chip. Therefore, the memory controller provides {an abstracted view of DRAM to the \om{CPU}}, preventing the programs \om{that run on the CPU} from directly interfacing with the DRAM \om{chip} \om{using low-level DRAM commands}. \om{However, experimental studies on DRAM typically require precise control of low-level DRAM commands (i.e., the ability to schedule any standard command at an arbitrary time). Thus, computer systems are limited in providing the level of control required by experimental studies over the DRAM interface to programs.}}}

\om{To overcome the limitations of existing computer systems, prior works develop DRAM testing infrastructures that allow issuing DRAM commands with {modified/}violated timing parameters to DRAM chips.}
{\om{Unfortunately,} {the two} existing FPGA-based open source DRAM testing infrastructures \om{(SoftMC~\cite{hassan2017softmc} and LiteX RowHammer Tester~\cite{litex.github})} suffer from critical issues that prevent researchers from {easily} {pursuing} the above research directions.} {These infrastructures} \One \om{\hh{do not fully expose the DRAM interface to test programs (i.e., {impose restrictions on the DRAM interface {that limit studies that can be performed}}), \Two are {fairly} difficult to use, or \Three are difficult to extend to support new DRAM interfaces}}.

\noindent
\changev{\ref{q:r1q1}}\tcadreva{\textbf{Interface Restrictions.}}
\om{Both SoftMC and Litex RowHammer Tester (LRT) expose a restricted view of the DRAM interface to test programs due to the limited capability provided by their instruction set architectures (ISA). First, SoftMC imposes restrictions on data patterns that can be used in initializing DRAM rows. Second, SoftMC {imposes} a non-deterministic delay in the order of microseconds between different parts of a DRAM test program\revdel{because SoftMC does not have loop or branch instructions in its ISA and test programs have to be transferred from the user's computer to the FPGA board in parts}. Third, LRT imposes \SI{10}{\nano\second} of delay between successive DRAM commands (i.e., a DRAM timing parameter can only be violated down to \SI{10}{\nano\second} and many standard DRAM timing parameters~\cite{jedec2017ddr4,jedec2012ddr3,jedec2020ddr5} \emph{cannot} even be violated). These restrictions \emph{limit} the types of experimental studies supported by both platforms.}

\noindent
\changev{\ref{q:r1q1}}\tcadreva{\textbf{Ease of Use.}}
\om{SoftMC and LRT are difficult to use. First, SoftMC \One is only prototyped on a discontinued FPGA board, and \Two has run-time dependencies {that} are obsolete and challenging to set up. Therefore, obtaining or setting up a system that readily supports SoftMC to conduct experimental studies is difficult. Second, LRT requires users to implement a new hardware module that post-processes experimental data read from DRAM (by using a hardware description language (HDL) such as Verilog) for every new type of experiment. Developing a hardware module is considerably difficult compared to developing a computer program for post-processing and requires expertise in HDL. As such, LRT \emph{cannot} be easily used as a general DRAM testing infrastructure.}

\noindent
\changev{\ref{q:r1q1}}\tcadreva{\textbf{Extensibility.}}
\hh{SoftMC \One support\atb{s} \emph{only} DDR3 and has a hardware design that is not \om{modular}, and \om{\Two has a {deprecated} development environment that consists of outdated hardware design tools.} Therefore\yct{,} it \emph{cannot} be easily modified to support DDR4 or other current and future DRAM interfaces.}

\noindent
\changev{\ref{q:r1q1}}\tcadreva{\textbf{Our goal}} {is to develop an infrastructure that is superior to past infrastructures by overcoming their shortcomings. To this end\tcadreva{, we design and develop \X{}}. \X{} \One \hh{fully exposes the DRAM interface to test programs through its nonrestrictive} \om{instruction set architecture,} \om{\Two provides easy-to-use C++ and Python programming interfaces, and open source working prototypes of itself on five different modern FPGA boards, \Three has a modular design and provides the necessary HW/SW components to facilitate {the} integration of new DRAM standards.} {\hh{Table~\ref{table:tools} presents} a qualitative comparison of \X{} against the two other open source DRAM testing infrastructures \hh{available today}.
\revdel{, we identify three major characteristics such an infrastructure should provide \tcadrevcommon{based on our analysis of prior open source DRAM testing infrastructures}\changev{\ref{q:r1q1}}\tcadreva{: i) \tcadrevcommon{no restrictions on the DRAM interface}, ii) \tcadrevcommon{easy to use}, and iii) \tcadrevcommon{extensible}}.} 
\revdel{\om{First}, the DRAM testing infrastructure \om{should} {not} restrict how a DRAM test interfaces with a DRAM chip but {should} {fully expose} the DRAM interface to a test program. \om{This is critical to provide users with the ability to perform any type of DRAM experiment.} {\om{Second}, the infrastructure \om{should} be easy-to-use. This is essential for the infrastructure to be widely used \om{by researchers and developers}. \om{Third}, the infrastructure \om{should} support a variety of the existing DRAM interfaces {(e.g., DDR3/4)} and \om{should} be easily {extensible} to support future DRAM interfaces. This is essential to enable research that studies the trends of DRAM scaling over a long time period as the DRAM interfaces evolve with the DRAM technology nodes.}}}

\noindent
\changev{\ref{q:r1q1}}\tcadreva{\textbf{Key Results.}}
\hh{We \om{\One}evaluate \X{}'s versatility in types of {experimental studies on DRAM} \tcadreva{(Section~\ref{sec:case-studies})}, \om{\Two}demonstrate \X{}'s ease of use \tcadreva{(Sections~\ref{sec:case-study-1} and \ref{sec:case-study-2})}, and \om{\Three}evaluate \X{}'s extensibility to new DRAM interfaces \tcadreva{(Section~\ref{sec:ddr3-modifications})}. First, \tcadreva{leveraging DRAM Bender's \emph{versatility} in conducting experimental studies on DRAM, we uncover two new insights into the RowHammer problem~\cite{kim2014flipping,mutlu2019retrospective} (Sections~\ref{sec:case-study-1} and~\ref{sec:case-study-2}) and demonstrate that processing-{using}-memory capability {(i.e., bulk bitwise execution capability)} exists in real DDR4 chips (Section~\ref{sec:case-study-3}). {Our results show that i) \hh{the effectiveness of} a double-sided RowHammer attack \hh{{at} causing bit-flips} highly depends on the \hh{order in which the two aggressor rows are activated and precharged}\tcadreva{, and ii)} data patterns supported by \X{} can uncover a larger set of bit-flips {in} a victim row compared to the data patterns commonly used by prior work.} Second, we demonstrate \X{}'s \emph{ease of use} by showing that it can be used to perform a RowHammer experiment in 12 lines of C++ code and a {bulk} bitwise AND/OR operation in just 3 lines of C++ code\om{, in working prototypes of DRAM Bender on five different modern FPGA boards}. Third, we demonstrate {that} \X{} is easy to extend (Section~\ref{sec:ddr3-modifications}). The required source code modifications to port \X{} to another FPGA board with a different DRAM interface are {small} (only 230 lines of additional Verilog and 30 lines of additional C++ code).\footnote{{On top of the initial implementation of \X{} with DDR4 support that comprises approximately 3400 lines of Verilog and 2000 lines of C++ code.}}}

\revdel{\om{First,} \hh{in Section~\ref{sec:case-studies},} we conduct {three} case studies to \hh{demonstrate the DRAM characterization capabilities of \X{} {and uncover two new insights into the RowHammer problem}.} We \hh{study} \om{\One}the effects of \hh{fine-grained DRAM} access patterns on RowHammer-induced bit-flips. Our results show that \hh{the effectiveness of} a double-sided RowHammer attack \hh{on causing bit-flips} highly depends on the \hh{order in which the two aggressor rows are activated and precharged}. {We study \om{\Two}the effects of random data patterns on RowHammer-induced bit-flips. {We show that the data patterns supported by \X{} can uncover a larger set of bit-flips {in} a victim row compared to the data patterns commonly used by prior work}.} {\om{Leveraging DRAM Bender's versatility in conducting experimental studies on DRAM},} {we demonstrate that processing-in-memory capability exists in DDR4 devices}. {W}e find that {the} {tested} DDR4 devices {can perform bitwise AND/OR operations in a proportion of the DRAM array}.} 

\revdel{\om{Second, we demonstrate the ease of use of \X{} in Sections~\ref{sec:case-study-1} and \ref{sec:case-study-2}. We show that \X{} can be used to perform a RowHammer experiment in 12 lines of C++ code and a bitwise AND/OR operation in just 3 lines of C++ code\om{, in working prototypes of DRAM Bender on five different modern FPGA boards}.}}
\revdel{\om{Third,} in Section~\ref{sec:ddr3-modifications}, we explain the changes required {to port \X{} to another DRAM board with a different DRAM interface.} We show that the required source code modifications are trivial and {support for a new {interface and a new} board} can be achieved with the addition of only 230 lines of Verilog and 30 lines of C++ code.}
\revdel{\footnote{{On top of the initial implementation of \X{} with DDR4 support that comprises approximately 3400 lines of Verilog and 2000 lines of C++ code.}}}}

\om{\X{} was first developed in 2019. Since then, it has been used in many studies~\cite{orosa2021deeper,kim2020revisiting,frigo2020trrespass,hassan2021utrr,olgun2021quactrng,koppula2019eden,yaglikci2022understanding,yaglikci2022hira,luo2023rowpress}. In this work, we aim to describe it in full, demonstrate its novel capabilities, and open source it so that it provides a useful research tool to the memory systems research and development community.}

\om{This paper} make\om{s} the following contributions:
\squishlist
    \item We develop \X{}, \om{an} FPGA-based open source DRAM testing \hh{infrastructure} that \om{overcomes the three major shortcomings of prior infrastructures. \X{} \One fully exposes the DRAM interface, \Two is easy to use, and \Three is easy to extend and modify.} {\X{} is the only \om{general} open source testing infrastructure to easily test DDR4 chips today.}
    \item \hh{We make {two} new observations {on RowHammer-induced bit-flips, a major reliability, safety, and security problem in modern DRAM chips.}} {These observations can be used to construct better RowHammer attacks and defenses.}
    \item We demonstrate that contemporary off-the-shelf DDR4 devices are capable of performing in-DRAM bitwise {Majority, AND, and OR} operations.
    \item We demonstrate the \emph{extensibility} of \X{} by \om{developing its prototypes for} five FPGA boards\hh{, which have either {DDR4 DIMM/SODIMM or DDR3 SODIMM} slots}.
    \item \om{We open source \X{}~\cite{self.github} to facilitate research on understanding the operation and characteristics of real DRAM chips. We also open source~\cite{utrr.github,quactrng.github} two prior works, U-TRR~\cite{hassan2021utrr} and QUAC-TRNG~\cite{olgun2021quactrng}, as example work{s} \X{} {has} enabled.}
\squishend

\section{Background and Motivation}
{We provide a concise background on DRAM organization, cell access, and timing parameters in this section. We refer the reader to many prior works for more detailed background on these subjects~\cite{liu2012raidr, liu2013experimental, keeth2001dram, mutlu2007stall, moscibroda2007memory, mutlu2008parbs, kim2010atlas, subramanian2014bliss, kim2014flipping, qureshi2015avatar, hassan2016chargecache, chang2016understanding, lee2017design,  chang2017understanding,  patel2017reaper,kim2018dram, kim2020revisiting, hassan2019crow, frigo2020trrespass, chang2014improving, chang2016low, ghose2018vampire, hassan2017softmc, khan2016parbor, khan2016case, khan2014efficacy, seshadri2015gather, seshadri2017ambit, kim2018solar, kim2019d, patel2019understanding, patel2020beer, lee2013tiered, lee2015decoupled, seshadri2013rowclone, luo2020clrdram, seshadri2019dram, wang2020figaro,orosa2021codic,wang2018reducing,ipek2008self,zhang2014half,kim2012case,olgun2022sectored,hassan2022case}.}

\subsection{DRAM Organization}
DRAM-based main memory is typically organized hierarchically, as depicted in \agy{F}igure~\ref{fig:dram-background}. A \emph{memory controller} in the CPU is connected to {a} \emph{DRAM module} over a \emph{DRAM channel}. Each module on a DRAM channel {comprises} multiple \emph{DRAM chips}, where \om{a} set of chips {that operate concurrently} is referred to as a \emph{DRAM rank}. A DRAM chip is partitioned into multiple \emph{DRAM banks} that operate independently. A \om{collection} of \emph{DRAM cells}, {\emph{row decoders}}, and {\emph{row buffers (sense amplifier arrays)}} constitute a DRAM bank. 

\begin{figure}[h]
    \centering
    \includegraphics[width=\linewidth]{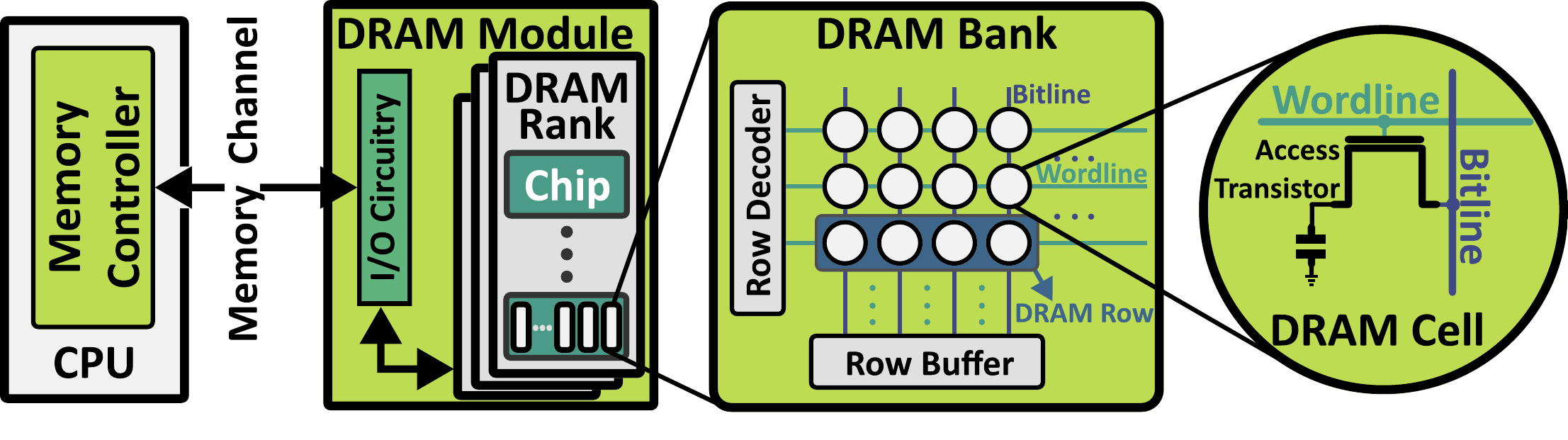}
    \caption{DRAM Organization}
    \label{fig:dram-background}
\end{figure}

DRAM cells that are laid out onto the same \emph{wordline} are accessed simultaneously, and referred to as a \emph{row}. Each cell in a row is connected to a different \emph{bitline} over an \emph{access transistor}. A bitline connects one DRAM cell from each row to a sense amplifier {in the row buffer}. A DRAM cell stores data as charge in a capacitor. A sense amplifier can read or manipulate a DRAM cell's value by sampling or driving a bitline.

\subsection{Accessing a DRAM Cell}
Initially, the bitline is \emph{precharged} to a {reference} voltage level ($V_{DD}/2$). The memory controller sends an \texttt{activate (ACT)} command to a \emph{row} to enable the wordline. This starts the \emph{charge-sharing} process, where the DRAM cell on the wordline shares its charge with its bitline. The sense amplifier samples the bitline voltage and detects a deviation towards $V_{DD}$ or $0$, depending on the value stored in the DRAM cell, and amplifies it. The DRAM cell can then be read or {its value updated} using \texttt{READ}/\texttt{WRITE} commands until the bank is finally \texttt{precharged (PRE)} to access another row.

\subsection{DRAM Timing Parameters}
Each DRAM standard defines a set of \emph{timing parameters}. {A timing parameter specifies the minimum time between two DRAM commands to guarantee correct operation.} We describe two relevant timing parameters:

\textbf{tRAS.} Time window between an \texttt{ACT} and the next \texttt{PRE} command. During \textbf{tRAS}, charge {of} the DRAM cells on the open DRAM row {is fully restored}.

\textbf{tRP.} Time window between a \texttt{PRE} and the next \texttt{ACT} command. \textbf{tRP} is required for the open wordline to be closed and the DRAM bitlines {be precharged} to $V_{DD}/2$.

\subsection{DDR PHY Interface (DFI)}

A memory controller is typically composed of two modules: \One the \emph{scheduler} generates and schedules the corresponding DRAM commands to serve memory requests (e.g., load and store requests), \Two the \emph{PHY layer} manipulates the physical interface signals to transmit DRAM commands and data over the DRAM interface to the DRAM {chips}. The DDR PHY Interface (DFI)~\cite{dfi} standardizes the communication between the memory controller and the PHY layer, enhancing the reusability of memory controller designs. 

DFI signals are broadly classified into \One \emph{control} signals that are used to issue DRAM commands, which are mostly analogous to the DRAM physical interface signals (e.g., \texttt{cas}, \texttt{ras}, and \texttt{we} signals), \Two \emph{read \& write data} signals that are used to transfer data over the DRAM interface. These signals in \om{the} DFI fully expose the underlying physical DRAM interface to the memory controller. We refer the reader to \om{the} DFI Specification~\cite{dfi} for more details.

\subsection{RowHammer}
{\lois{RowHammer}, first introduced in 2014~\cite{kim2014flipping}, is a circuit-level phenomen\om{on} in DRAM devices, where frequent activation of an \emph{aggressor} DRAM row induces bit-flips on DRAM cells in nearby \emph{victim} rows. \lois{RowHammer} is an important DRAM {security/reliability/safety} vulnerability that becomes more severe as DRAM devices scale {to smaller technology nodes}~\cite{kim2014flipping, kim2020revisiting,orosa2021deeper}. A \emph{\lois{RowHammer} attack} can flip critical bits in memory and can be used to {escalate privilege and break confidentiality} in real systems ({as shown in, e.g.,{~\exploitingRowHammerAllCitations{}}).}

\subsection{{\om{FPGA-based} DRAM Testing Infrastructures}}
\label{sec:existing-infrastructures}

\om{SoftMC~\cite{hassan2017softmc,softmcgithub} and Litex RowHammer Tester (LRT)~\cite{litex.github} are the only two open source FPGA-based DRAM testing infrastructures that exist today.}

\noindent
\textbf{SoftMC~\cite{hassan2017softmc,softmcgithub}} is an open source DDR3 testing infrastructure. SoftMC issues DDR3 command traces to the DDR3 device under test. A host machine transmits these command traces to the FPGA board that implements SoftMC. 

\noindent
\textbf{Litex RowHammer Tester\cite{litex.github}} is a RowHammer testing infrastructure that is primarily designed to conduct \lois{RowHammer} experiments on \yct{DDR3/4 and LPDDR4} chips. LRT receives DRAM command sequences from the host machine and issues them to the DRAM device.

\om{Both SoftMC and LRT suffer from multiple shortcomings that make them inadequate to {satisfy} the requirements of a versatile, easy-to-use, and extensible FPGA-based DRAM testing infrastructure. We describe these shortcomings in detail:}

\noindent
{\textbf{Shortcomings of SoftMC.} \om{First}, SoftMC imposes restrictions on DDR3 data transfers. By replicating an 8-bit data pattern provided by the user, SoftMC initializes DRAM cache blocks (512-bits) with repetitive data. This prevents its users from exploring the large search space of data patterns that can be used in experimental DRAM studies (Section~\ref{sec:case-study-2}). Second, due to its DDR3 command trace-based execution model, SoftMC {imposes} a non-deterministic delay on the order of a few microseconds between DDR3 command traces that are executed back-to-back. This prevents a user from running a class of experiments (e.g., energy measurements, Section~\ref{sec:power-consumption-studies}) {that} require precise command timings over a long period of time (i.e., on the order of seconds or minutes, where SoftMC {needs to execute} multiple DRAM command sequences {and induces multiple non-deterministic delays on the order of microseconds}). \om{Third}, SoftMC is {fairly} difficult to use and extend. {It is only prototyped on the ML605 FPGA board~\cite{ml605}, which is no longer manufactured.} {{SoftMC's} development environment {contains} obsolete and discontinued software and hardware components~\cite{xilinxise,jacobsen2015riffa}.}}
\om{Fourth, SoftMC's DDR3 interface implementation is not decoupled from its remaining hardware via a well-defined interface (i.e., {SoftMC is} hard-coded to work with DDR3 modules). Thus, implementing new DRAM standards (e.g., DDR4) require{s} major {intrusive} modifications to {SoftMC's} hardware {design}.}

\noindent
{\textbf{Shortcomings of LRT.} First, LRT does not support timing delays between successive DRAM commands smaller than \SI{10}{ns}~\cite{lrt10ns}. This greatly limits the DRAM experiment space supported by LRT. For example, many prior works that evaluate DRAM under reduced activation latency~\cite{kim2019d,gao2019computedram,kim2018solar,kim2018dram,chang2016understanding} and under greatly reduced $tRAS$-$tRP$~\cite{gao2019computedram,olgun2021quactrng} could not conduct their experiments using LRT. {Second, LRT can transfer data between the host machine and the FPGA board at only up to 1.9 Mb/s~\cite{lrt-issue}. A slow communication channel between the host machine and the FPGA board can greatly slow down experiments, as DRAM experiments often require a large amount of data to be transferred to the host machine for analysis (Section~\ref{sec:hardware-ip-cores}).} To alleviate the overheads of data transfer and enable practical testing times for RowHammer experiments, LRT implements a specialized hardware module for post-processing. However, it would be impractical to develop a specialized module for each type of DRAM experiment\om{. Thus LRT is difficult to use as a general DRAM testing platform}.}

\subsection{Motivation {for \X{}}}
\label{sec:motivation}

{None of the existing open source DRAM testing infrastructures can be used to easily and comprehensively test state-of-the-art, DDR4 DRAM chips.} \om{Therefore,} it is essential to develop an open source, versatile, easy-to-use, \om{and extensible} DRAM testing infrastructure to enable {experimental studies on new DRAM chips}. Based on this requirement and our evaluation of \hh{the} existing DRAM \hh{testing} infrastructures, we define \hh{three} key features that a DRAM \hh{testing} infrastructure should provide. 

\noindent
\textbf{Nonrestrictive Interface.} \hh{A DRAM interface typically provides} commands that can be used to access and manipulate DRAM cells (e.g., \texttt{READ} and \texttt{WRITE}), DRAM bank state (e.g., \texttt{ACT} a row), and device state (e.g., \hh{enter/exit} self-refresh mode). To enable a rich DRAM {testing environment}, an infrastructure must provide a \emph{nonrestrictive} application programmer interface (API) that \emph{directly exposes} the DRAM interface to \hh{a test program}. The API must not enforce any timing constraints nor any restrictions on the ordering of DRAM commands issued to the DRAM device.

\Copy{R2/4z}{
\changev{\ref{q:r2q4}}
\noindent
\tcadrevb{\textbf{Ease of Use.} The infrastructure must be easy to \One develop, for \tcadrevcommon{low-effort} integration of new DRAM interfaces and prototyping on new FPGA boards, \Two {build}, to easily bring up a working prototype using software and hardware tools that are widely available, and \Three use, to seamlessly design and conduct DRAM experiments.}}

\noindent
\textbf{Extensibility.} {DRAM interfaces continue to evolve as {new workloads emerge and DRAM technology scales}. Therefore, a DRAM testing infrastructure {must} be easily {extensible} to support the new DRAM interfaces {(e.g., HBM2~\cite{jedec2015hbm}, DDR5~\cite{jedec2020ddr5})} to perform \tcadrevcommon{experiments} on the latest technology DRAM devices.} 

{E}xisting DRAM testing infrastructures \hh{lack at least one of these features {and cannot be used to easily test state-of-the-art DDR4 chips}. Thus\yct{,} there is a need for an infrastructure that provides all three features to stimulate future DRAM research}.

\section{\X}

\X{} is an extensible, \om{versatile,} and easy-to-use FPGA-based DRAM testing infrastructure. \om{It} comprises key hardware components that facilitate the integration of new DRAM interfaces in a modular design and provides a \atb{nonrestrictive} \om{and easy-to-use} application programmer interface (API) that directly exposes the DRAM interface to the user. 

Figure~\ref{fig:block-diagram} depicts \hluo{the} block diagram of \X{}. A\atb{n} \hluo{FPGA board that implements \X{}} is attached \hluo{to the host machine} over a high-speed interface (e.g., PCIe\ext{~\cite{pcie}}). 
\atb{A} key component of \X{} is the \emph{programmable core}, where the DRAM tests, organized as self-contained \X{} programs, are executed. The programmable core implements the \X{} ISA{, {which} exposes} key functionalities to form a DRAM test program. 

\begin{figure}[!ht]
    \centering
    \includegraphics[width=\linewidth]{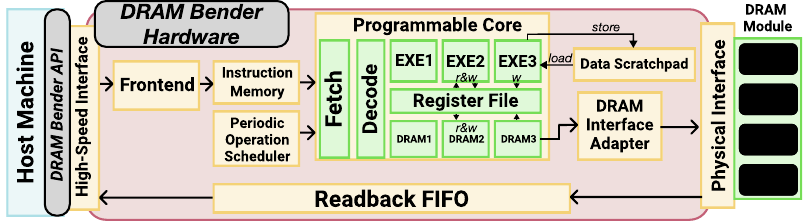}
    \caption{Block diagram of \X{}}
    \label{fig:block-diagram}
    \vspace{-2mm}
\end{figure}

\subsection{\X{} ISA}

\label{sec:x-isa}
\hluo{The \X{} ISA} (Table~\ref{table:isa-description}) exposes key \hluo{functionalities} that \hluo{the user needs to write a DRAM test program}. \hluo{{For example, these functionalities include} issuing DRAM command sequences with control flow (e.g., in a loop)} and dynamically modifying the wide (e.g., 512-bits) data word that is written to DRAM with the execution of a \texttt{WRITE} command. \hluo{There are two types of instructions in the \X{} ISA: 1) RISC-like regular instructions that operate on the registers of the programmable core (e.g., arithmetic instructions and control-flow instructions), and 2) DRAM instructions that send DRAM commands to the DRAM module under test.} A \emph{DRAM instruction} is composed of four\footnote{Most FPGA DRAM PHY {designs} serialize multiple DRAM commands that they receive each cycle and issue the commands to the DRAM device one by one at a higher frequency. {The PHY design we use (Section~\ref{sec:hardware-ip-cores}) can be configured to receive up to four DRAM commands each cycle.} By executing four DRAM commands per cycle, \X{} can be clocked at a four times slower frequency than the minimum clock frequency defined by a DRAM standard.} arbitrary DRAM commands. To fully expose the DRAM interface to the user, \X{} permits arbitrary ordering of DRAM commands in a DRAM instruction.

\begin{table}[h]
\centering
\renewcommand{\arraystretch}{1.1}
\footnotesize
\captionsetup{justification=centering}
\caption{DRAM Bender ISA \omcr{d}escription}
\begin{tabular}{cll}
\hline
\textbf{Type} & \textbf{Instruction} & \textbf{Description} \\ \hline \hline
\multirow{2}{*}{\rotatebox[origin=c]{90}{MEM}}     & LD    & Load one word from the data scratchpad \\
                            & ST    & Store one word to the data scratchpad \\ \hline
\multirow{5}{*}{\rotatebox[origin=c]{90}{ARITH.}} & AND/OR/XOR   & Bitwise AND/OR/XOR two registers \\
                            & ADD/SUB & Add/Sub two registers\\
                            & MV    & Move a register's value to another register \\
                            & SRC   & Cyclic shift a register to right \\
                            & LI    & Load a register with an immediate value \\ \hline
\multirow{3}{*}{\rotatebox[origin=c]{90}{CNTRL.}}   & BL/BEQ    & Branch if less than/equal \\
                                & JUMP  & Jump to an address \\
                                & SLEEP & Stall for a number of cycles \\ \hline
\multirow{4}{*}{\rotatebox[origin=c]{90}{MISC.}}      & LDWD  & Load the wide data register \\
                            & LDPC  & Read a performance counter \\
                            & SRE/SRX & Self-refresh entry/exit \\
                            & END   & Stop executing the program \\ \hline

\end{tabular}
\label{table:isa-description}
\end{table}

\Copy{R3/2b}{
\noindent
\textbf{Instruction Encoding.} Figure~\ref{fig:isa-encoding} depicts the encoding of a subset of \X{} ISA instructions. A DRAM command is encoded in 18 bits. The first two fields in a DRAM command specif\om{y} the registers that store the address arguments of that DRAM command. For example, \om{the first field \om{of an ACT command}} indicates the \emph{bank address} register and the second field indicates the \emph{row address} register. {The DRAM command encoding size determines the instruction size}, and it is 72 bits.\footnote{$18\times{}4=72$, as there are four DRAM commands in a DRAM instruction.} We reserve 2 bits in \om{the} DRAM command encoding's op-code field to implement support for new commands in emerging and future DRAM interfaces (e.g., MPC in DDR5~\cite{jedec2020ddr5}). 
\changev{\ref{q:r3q2}}\tcadrevc{We present a step-by-step guide on extending the ISA with new instructions on \X{}'s Github repository~\cite{self.github}.}}

\begin{figure}[h]
    \centering
    \includegraphics[width=.49\textwidth]{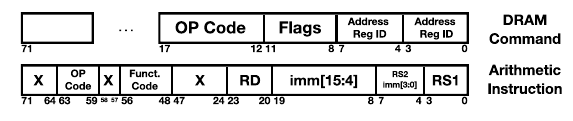}
    \caption{\X{} {instruction} encodings}
    \label{fig:isa-encoding}
\end{figure}

\noindent
\textbf{Registers.}
\X{} implements 13 general purpose registers (GPRs) \hluo{and four special registers.} The first three \hluo{special} registers are called \emph{address stride registers}. These registers \hluo{determine by how much the \emph{address registers} are automatically incremented with the execution of a DRAM instruction when} the corresponding bit in the \emph{flags} field is set. Automatically incrementing address registers allows \hluo{\X{} test programs to} support a wide range of access patterns. {For example, the user can tightly schedule \texttt{ACT} commands targeting different DRAM banks by automatically incrementing the bank address register value with each \texttt{ACT} command.}
The \hluo{fourth} special register is the \emph{wide-data register}. This register is as large as the transfer size of the DRAM interface (e.g., 512 bit\yct{s} for an 8-chip DDR4 module~\cite{jedec2017ddr4}). {The wide-data register} can only be manipulated using the \texttt{LDWD} instruction, and is \hluo{used} by the \texttt{WRITE} command. The wide-data register allows users to initialize DRAM devices with arbitrary data patterns at the data transfer granularity (e.g., 512 bits in a DDR3/4 module). In contrast, a prior testing \atb{infrastructure}~\cite{hassan2017softmc} allows the users to \emph{only} specify 8-bit data patterns, \hluo{which {are} replicated 64 times to form the 512-bit transfer data, {significantly} limiting} the data patterns that can be used in DRAM experiments \hluo{(e.g., {the prior testing infrastructure \emph{cannot} randomly initialize DRAM chips}).}

\subsection{\X{} Hardware Modules}

\hluo{The hardware components of \X{} are:} \One a specialized programmable core that implements the \X{} ISA, \Two a modular memory controller subsystem, and \Three multiple memory elements that are used to store the \X{} program and data, and buffer the data read back from the DRAM device under test.

\subsubsection{Programmable Core}
{\X{} comprises a programmable core that implements the \X{} ISA}. To {enable} the programmable core to achieve high frequencies that are required to interface with DRAM devices (\hluo{under normal operating conditions, \om{the minimum operating frequency for DDR4 devices is} 666 MHz~\cite{jedec2017ddr4}}), we design it as an in-order core with five pipeline stages: \One \emph{fetch}, \Two \emph{decode} and \Three three \emph{execute} stages. The fetch stage fetches a single instruction (72 bits) from the \emph{instruction memory}. The decode stage decodes the instruction into either an \emph{execute} {micro-operation ($\mu$-op)} or into four \emph{DRAM} $\mu$-ops.\footnote{{\X{} needs to execute four DRAM {commands} to commit an instruction that encodes DRAM commands. To {distinguish between} what is executed by \X{}'s hardware (i.e., the four DRAM {commands}) and the instruction execution capability exposed to the user via \X{}'s ISA (i.e., the instruction that encodes four DRAM commands), we refer to what \X{}'s hardware executes as micro-operations ($\mu$-ops). {\X{} decodes a \X{} ISA instruction into one (in case of a regular instruction) or four $\mu$-ops (in case of a DRAM instruction). A $\mu$-op signal flows through pipeline stages with the necessary information (e.g., execution flags, register identifiers, and operands) to correctly execute its corresponding \X{} ISA instruction.}}} The execute stages are separated into two pipelines: \One \yct{\emph{execute}} pipeline, which executes regular $\mu$-ops, and \Two \emph{DRAM} pipeline, which executes four DRAM $\mu$-ops \hluo{per} cycle. 

\noindent
\textbf{DRAM $\mu$-ops.} Each DRAM $\mu$-op corresponds to a DRAM command. {To execute a DRAM $\mu$-op,} \X{} first reads the GPRs to obtain the address operands (e.g., bank address of a \texttt{PRE} command) of {the corresponding DRAM command} and then issues the command to the DRAM device over the physical interface. The timing parameters of a DRAM command sequence are indirectly specified by either \hluo{regular} \emph{SLEEP} instructions or DRAM \emph{NOP} $\mu$-ops. 

\noindent
\textbf{Pipeline Dependencies.} Both pipelines read from and write to the register file at the second execute stage to maintain a simple design. The only exception to this is \emph{LOAD} instructions\hluo{, which} write to the register file at the third execute stage \hluo{for} better design timing closure during implementation. Control-flow instructions (i.e., branches) simply stall the pipeline until they are resolved at the third execute stage. Since the penalty of control-flow instructions is deterministic (six clock cycles), stalling the pipeline does not affect the user-specified timing parameters of the DRAM command sequences \om{in an unpredictable way}. 

\subsubsection{DRAM Interface Adapter}
\label{sec:dram-interface-adapter}
\X{} implements the DRAM interface adapter, which acts as an intermediary between the DRAM pipeline and the physical DRAM interface. The DRAM pipeline simply outputs a \emph{one-hot encoded signal} for each DRAM command in a DRAM instruction. For example, a DRAM instruction with an \texttt{ACT} as its first and a \texttt{WRITE} as its third command would set the least-significant bit of the \texttt{dram\_act} signal to ``logic-1'' and the third least-significant bit of the \texttt{dram\_write} signal to ``logic-1'', while leaving all other signals at ``logic-0''. These signals are picked up by the \emph{DRAM adapter} module, which essentially translates from this simplified interface (one-hot encoded signals) to a more complex PHY interface. The adapter is decoupled from the rest of the design such that it can be easily modified to integrate new DRAM interfaces. 

Developers can easily extend \X{} to support another DRAM interface by \One defining new one-hot encoded signals that are set with the execution of new DRAM commands introduced in the new interface and \Two modifying the adapter to match with the new DRAM PHY interface.

\subsubsection{Memory Elements}
\label{sec:memory-elements} 
\X{} comprises multiple memory elements that store programs, program data\tcadrevcommon{,} and data read back from DRAM during an experiment. The \emph{instruction memory} stores the \X{} program and it is initialized by the \emph{frontend} module, which receives data over the high-speed interface (e.g., PCIe) from a host machine. The \hluo{\emph{data scratchpad}} stores program data and it can be accessed and manipulated using \texttt{LD} and \texttt{ST} instructions. 

\subsubsection{Periodic Operation Scheduler} To maintain correct operation on the DRAM interface, DRAM devices typically need to receive a set of commands periodically.\footnote{For example, to compensate for voltage and temperature changes (ZQ Calibration~\cite{jedec2017ddr4}) \om{and} refresh leaky DRAM cells (DRAM Refresh).} The \emph{periodic operations scheduler (POS)} performs these operations either with predefined (periodic \texttt{READ}, \texttt{ZQS}) or configurable (periodic refresh) periods. POS stores three \X{} programs \om{in its memory to} \om{conduct periodic \texttt{READ}, \texttt{ZQS} calibration, and refresh operations}. It arbitrates between these programs to correctly schedule the required periodic operation to the DRAM device.

\noindent
\subsubsection{Readback FIFO} 
Typically, there is {a} discrepancy between the \hluo{bandwidth of the high-speed host-FPGA interface and the DRAM interface on FPGA boards because the DRAM interface can be clocked at higher speeds.} To alleviate \hluo{the negative effects} (e.g., pipeline stalls) {due to} this discrepancy on experiments conducted using \X{}, we buffer the data read back from the DRAM device in a FIFO. The \emph{readback FIFO} buffers data read from the DRAM device as it is sent to the host machine over the high-speed interface.

\noindent
\textbf{{Avoiding} Readback {Stalls}.} The readback FIFO may fill up during the execution of a \X{} program, in which case {the host machine must read} the FIFO to allow the upcoming \texttt{READ} commands to execute. \X{} adopts a preemptive mechanism to stall programs at a timing-non-critical point (i.e., while executing regular instructions) to prevent the user-specified timing parameters of DRAM command sequences from being violated. {In other words, \X{} \emph{never} stalls a program while executing a continuous sequence of DRAM commands.} 

\X{} stalls a program prior to the execution of a {continuous} DRAM command sequence {that} contains \texttt{READ} commands {(DRAM command sequences without a \texttt{READ} command do \emph{not} cause readback {stalls})}. The API automatically inserts {into \X{} programs} hints which specify the {number} of \texttt{READ} commands that will be executed in the next continuous {DRAM command sequence}. \X{} {lets} such {command sequences} execute \emph{only} if there is enough space within the readback FIFO, and stall\omcr{s} them otherwise.

\subsection{\X{} Software Modules}

\hluo{\X{} contains two major software modules:} \One the application programmer interface, and \Two program debugger.

\subsubsection{Application Programmer Interface (API)}

\X{} provides an easy-to-use and \atb{nonrestrictive} application programmer interface (API) implemented in C++. \hluo{This} API is \emph{extensible} \hluo{with} a modular design that allows developers to easily integrate new DRAM interfaces to \X{}, and adapt \X{} to new FPGA boards. The API is composed of three components: \One \hluo{the \texttt{program} class} defines how \X{} programs are created, \Two \hluo{the \texttt{platform} class} exposes the FPGA-based platform to the user via key functions, \Three \hluo{the \texttt{board} class} implements wrapper functions that communicate with the FPGA board over that FPGA board's interface (e.g., PCIe drivers). {\X{} also implements a \texttt{Python} interface that provides the same functionality as the C++ API using \texttt{pybind11}~\cite{pybind11.github}. The \texttt{Python} interface allows 1) agile development and verification of \X{} programs, and 2) simple workflow for easy integration of popular data analysis and visualization frameworks (e.g.,~\cite{reback2020pandas, Waskom2021seaborn}).} We describe the {two} components in the API in more detail{.}

\noindent
\textbf{Program.} The \texttt{program} class encapsulates functions that are used to create \X{} programs. A \X{} program is internally represented as a list of instructions by the API. The user constructs this list by appending instructions to {the list} one by one.
Users can \hluo{also} append labels to a program which \hluo{can be} used as branch targets of control-flow instructions. This convention allows the user to easily create \X{} programs.
To demonstrate the ease of use, {L}isting~\ref{list:read-test-code} depicts a {C++ code example} that uses the \X{} API to create a \X{} program that reads from a DRAM row. {Our open source repository~\cite{self.github} contains other C++ code examples (e.g., for characterizing data retention time) that the users can benefit from or build upon.}

\vspace{2mm}
\begin{lstlisting}[caption={Example {\X{}} program that reads a row},captionpos=b,label={list:read-test-code}]
// The program |{below}| reads all cells in Bank 0, Row 0
// R5: bank address, R4: row address, R3: column address
// CASR: column stride register, R6: last column to read
Program p; // A sequentially-generated instruction list
p.appendLI(R5, 0); // Add a new instruction to the list
p.appendLI(R4, 0);
p.appendLI(R3, 0);
p.appendLI(CASR, 8); // Column address will increase by 8
p.appendLI(R6, 1024); // Read until 1024th column address
p.appendACT(R5, false, R4, false, |\textcolor{red}{11}|);
p.appendLabel(|\textcolor{violet}{"read"}|);
p.appendREAD(R5, false, R3, true, false, false, |\textcolor{red}{0}|);
p.appendBL(|\textcolor{violet}{"read"}|, R3, R6);
p.appendPRE(R5, false, false, |\textcolor{red}{0}|);
\end{lstlisting}
\vspace{3mm}

The \hluo{\texttt{program} class defines} a set of functions that are based on the template \texttt{append<Instruction>}. Each \texttt{append} function either inserts a regular instruction or a DRAM command {into} the program. \texttt{append} functions that insert regular instructions have RISC-like argument orderings, e.g., {\texttt{appendLI(R5, 0)} (line {5})} inserts a {load immediate} instruction that {sets the value of} R5 to 0. The arguments of \texttt{append} functions that insert DRAM commands into the program {(lines {10, 12, and 14})} are ordered as \One address register identifier {(one or two)}, \Two address register increment flag {(one or two)}, \Three other flags, and \Four an optional delay (highlighted in red in Listing~\ref{list:read-test-code}).

Using the optional delay, users can easily create DRAM command sequences with arbitrary timing parameters. The address register increment flag allows for access patterns with arbitrary strides. As an example, the arguments of the \texttt{appendACT} function \hluo{(line {10})} in listing~\ref{list:read-test-code} are \One \textbf{R5}, bank address register (BAR), \Two \textbf{false}, \emph{do not} increment BAR by bank address stride register, \Three \hluo{\textbf{R4}}, row address register (RAR), \Four \textbf{false}, \emph{do not} increment RAR by row address stride register, and \Five wait for \textbf{11} additional DRAM command bus cycles after issuing the \texttt{ACT} command, in order from left to right. In the \texttt{appendREAD} function, the user reads all DRAM cache blocks in a DRAM row by setting the column address register increment flag, without having to execute \texttt{ADD} instructions {inbetween} \texttt{READ} command{s}. This allows the user to schedule DRAM commands with as tight timing delays as possible.

\noindent
\textbf{Platform.} \hluo{The \texttt{platform} class defines methods to} \One execute programs on the \om{infrastructure}, \Two manipulate the state of the platform (e.g., reset, enable auto-refresh), and \Three retrieve data from the platform over the high-speed interface.

\noindent
\textbf{Board.} The {\texttt{board}} class defines how \hluo{data is transferred} between the host machine and the FPGA \om{infrastructure} over the high-speed interface. It exposes two key functions to the platform class{,} \texttt{sendData} and \texttt{receiveData}{,} which are overloaded by the code that implements the FPGA board-specific interface (e.g., Xilinx XDMA~\cite{xdma}). \texttt{sendData} is used by the platform \om{class} to send a program to the FPGA, and \texttt{receiveData} is used to read data {back} from the DRAM device under test.

\subsubsection{{Program} Debugger}
\label{sec:debugger}
\Copy{R2/4}{\X{} provides an easy-to-use program debugger. The debugger is used to analyze and debug \X{} programs generated by the API without access to the hardware setup. It uses the Vivado Simulator~\cite{vivadosimulator} to run timing-critical simulations of the hardware setup, which includes a DRAM device model. Using the DRAM model, the debugger can inform the user {of} every timing violation, which {enables} the user {to} easily understand if their program runs correctly (i.e., it does not violate {DRAM} timing parameters unexpectedly). \changev{\ref{q:r2q4}}\tcadrevb{The tight integration of the debugger and the Vivado Simulator allows users to view waveform diagrams and select which hardware signals to debug conveniently, using Vivado's graphical user interface.}}

\del{\X{}'s debugger provides a similar functionality as standard software debuggers do. It supports placing breakpoints in a \X{} program, advancing the program counter by one or many instructions, and probing the content of the register file and the scratchpad memory.}

\subsection{\X{} Prototypes}
\label{sec:prototypes}
We implement \X{} on five different FPGA boards that support different DRAM standards (DDR3, DDR4) and DRAM modules with different form factors (SODIMM, RDIMM, UDIMM), to demonstrate the extensibility of \X{}. {Among these five prototypes, we highlight the Xilinx Alveo U200 board prototype. When augmented with a SODIMM to DIMM adapter (e.g., JET-5608AA~\cite{jet5608}), this prototype can be used to test a large majority of all available DRAM modules.\footnote{{More details are in our GitHub repository~\cite{self.github} including a pictorial depiction of how the SODIMM to DIMM adapter is used with the Xilinx Alveo U200 board prototype to test SODIMMs.}}}
Different modules can contain DRAM chips built using different packaging technologies (e.g., FBGA 78 and FBGA 96). A \X{} user can find out what packaging technologies are used in chips on a DRAM module by inspecting the relevant module manufacturer specifications. Therefore, a DRAM Bender user can test different DRAM packages. Table~\ref{table:prototypes-table} describes the details of \X{} prototypes on different boards. \changev{\ref{q:r3q1}}\tcadrevc{Table~\ref{table:fpga-utilization} shows the FPGA resource utilization for each \X{} prototype.}

In all prototypes, the \emph{instruction memory} stores \emph{2048 instructions}, the \emph{scratchpad} stores \emph{1024 words (32-bits)}, and the \emph{read-back FIFO} is \emph{32 KiB} ({i.e.,} buffers up to 512 DRAM data transfers). {{\X{}} support{s} a DRAM timing resolution (i.e., {the} minimum distance between two consecutive DRAM commands) of \SI{1.5}{\nano\second} {in DDR4 boards} and {a DRAM timing resolution of \SI{2.5}{\nano\second} in the DDR3 board}.}\footnote{{We configure our prototypes such that the DRAM interface operates at the minimum standard frequency. This allows for \X{}'s hardware to satisfy the tight FPGA timing constraints in a subset of the FPGA boards (Bittware XUSP3S and Xilinx ZC706) with fairly old and resource-bound FPGAs. When necessary, \X{} can be implemented on other FPGA boards with newer FPGAs to satisfy a finer timing resolution out of the box (e.g., at least as fine as \SI{1.25}{\nano\second} in {Bittware} XUPVVH~\cite{XUPVVH}).}}

\begin{table}[h]
\footnotesize
\caption{\X{} FPGA prototypes}
\begin{tabular}{lll}
\textbf{FPGA Board}        & \textbf{DRAM Standard} & \textbf{Module Form Factor} \\ \hline \hline
Xilinx Alveo U200~\cite{Alveo} & DDR4          & {[U/R/SO]DIMM}   \\
Bittware XUSP3S~\cite{XUSP3S}   & DDR4          & SmaODIMM             \\
Bittware XUPP3R~\cite{XUPP3R}   & DDR4          & RDIMM              \\
Bittware XUPVVH~\cite{XUPVVH}   & DDR4          & RDIMM              \\
Xilinx ZC706~\cite{zc706}     & DDR3          & SODIMM            
\end{tabular}
\label{table:prototypes-table}
\end{table}

\begin{table}[h]
\Copy{R3/1table}{
\footnotesize
\caption{\tcadrevc{\X{} FPGA resource utilization}}
\begin{tabular}{l|ccccc}
\textbf{Resource}        & \textbf{XUSP3S} & \textbf{XUPP3R} &\textbf{XUPVVH} &\textbf{U200} &\textbf{ZC706} \\ \hline \hline
LUT&6.59\%&3.25\%&2.94\%&3.01\%&7.90\%\\
LUTRAM&2.55\%&0.34\%&0.34\%&0.34\%&1.68\%\\
FF&3.46\%&1.80\%&1.69\%&1.78\%&3.08\%\\
BRAM&5.21\%&4.75\%&4.99\%&4.61\%&8.17\%\\
DSP&0.39\%&0.04\%&0.03\%&0.04\%&0.00\%\\
IO&17.09\%&17.09\%&23.24\%&18.34\%&32.60\%\\
GT&12.50\%&10.53\%&8.33\%&33.33\%&25.00\%\\
BUFG&1.25\%&1.17\%&1.98\%&1.11\%&28.13\%\\
MMCM&6.25\%&3.33\%&8.33\%&3.33\%&25.00\%\\
PLL&9.38\%&5.00\%&12.50\%&5.00\%&12.50\%\\
PCIe&25.00\%&16.67\%&16.67\%&16.67\%&100.00\%\\
\end{tabular}
\label{table:fpga-utilization}
}
\end{table}

\noindent
\textbf{Controlled Environment.} To conduct experiments in a controlled environment, \lois{our infrastructure incorporates a} programmable \lois{Maxwell FT200}~\cite{maxwellFT200} temperature \lois{controller (Figure~\ref{fig:infraestructure}a) connected to silicone rubber heaters {attached} to both sides of the DRAM module (Figure~\ref{fig:infraestructure}b)}. \lois{To measure the actual temperature of DRAM chips, we use a thermocouple, which we place between the rubber heaters and the DRAM chips.  We connect the heater pads and the thermocouple to the temperature controller. \Copy{R1/4}{The temperature controller keeps the temperature stable via implementing a closed-loop PID controller.\footnote{{In our experiments, we typically use the temperature range of [\SI{30}{\celsius}, \SI{85}{\celsius}]. Based on the data sheets~\cite{maxwellFT200,thermocouple} of the temperature controller and the sensor, we expect the maximum error in temperature measurements to be within $\pm$\SI{1.11}{\celsius}.}}} \del{The host machine communicates with the temperature controller via an RS485 channel. In our tests using this infrastructure, we measure temperature with an accuracy of $\pm$\SI{0.1}{\celsius}.}}

\begin{figure}[h]
    \centering
    \includegraphics[width=.49\textwidth]{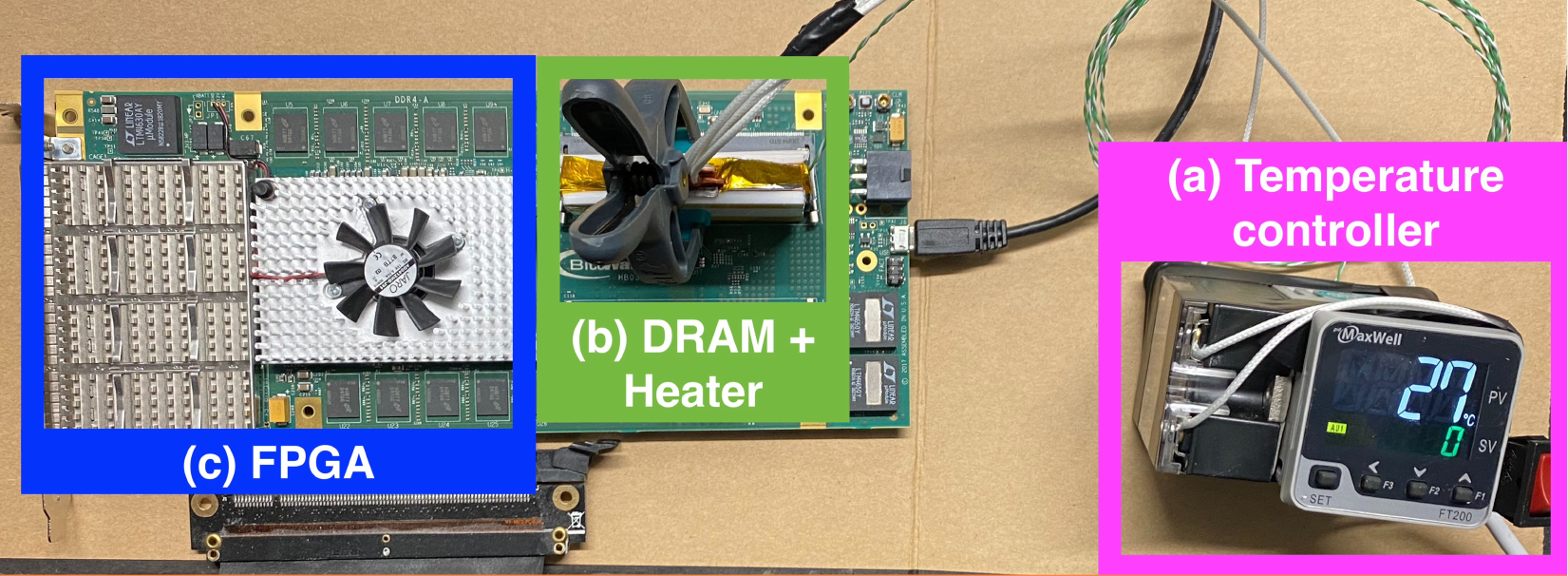}
    \caption{\lois{\X{} Hardware Infrastructure}}
    \label{fig:infraestructure}
\end{figure}

\noindent
\textbf{Temperature Measurement Stability.} We conduct a temperature controller stability experiment to demonstrate that the temperature we measure using our setup is stable and remains close to the target temperature throughout a day. First, we set the target temperature to \SI{80.0}{\celsius} in three {different} {prototypes (one of which is} shown in Figure~\ref{fig:infraestructure}).\footnote{{We use one Bittware XUSP3S (Prototype 1) and two Xilinx Alveo U200 (Prototype 2 and 3) FPGA prototypes.}} Second, we wait for the temperature to stabilize. Third, we record the temperature measurements every 5 seconds for 24 hours. We observe that the maximum variation in temperature is \SI{0.2}{\celsius} (i.e., the highest recorded temperature is \SI{80.2}{\celsius} and the lowest recorded temperature is \SI{79.8}{\celsius}). Table~\ref{table:temperature-stability} shows the proportion of temperature measurements that correspond to \SI{80.2}{\celsius}, \SI{80.1}{\celsius}, \SI{80.0}{\celsius}, \SI{79.9}{\celsius}, and \SI{79.8}{\celsius} among all measurements.

\newcolumntype{P}[1]{>{\centering\arraybackslash}p{#1}}
\begin{table}[h]
\caption{{Observed temperature measurements when the target temperature is set to \SI{80.0}{\celsius} in three different prototypes.}}
\centering
\footnotesize
\begin{tabular}{|c||P{1.5cm}|P{1.5cm}|P{1.5cm}|}
\hline
\multirow{2}{*}{\shortstack{{Measured}\\{Temperature ($^{\circ}$C)}}} & \multicolumn{3}{c|}{{Proportion of measurements} (\%)}  \\ \cline{2-4} 
                                 & {Prototype 1} & {Prototype 2} & {Prototype 3} \\ \hline\hline
80.2                             & 0.07    & 0.02    & 0.01    \\ \hline
80.1                             & 11.73   & 5.20    & 7.58    \\ \hline
{80.0}                             & 77.34   & 88.96   & 86.28   \\ \hline
79.9                             & 10.13   & 5.81    & 6.08    \\ \hline
79.8                             & 0.74    & 0.02    & 0.05    \\ \hline
\end{tabular}
\label{table:temperature-stability}
\end{table}

{Figure~\ref{fig:temperature-stability} depicts the measurements pictorially. We conclude that our infrastructure enables stable temperature control.}

\begin{figure}[h]
    \centering
    \includegraphics[width=\linewidth]{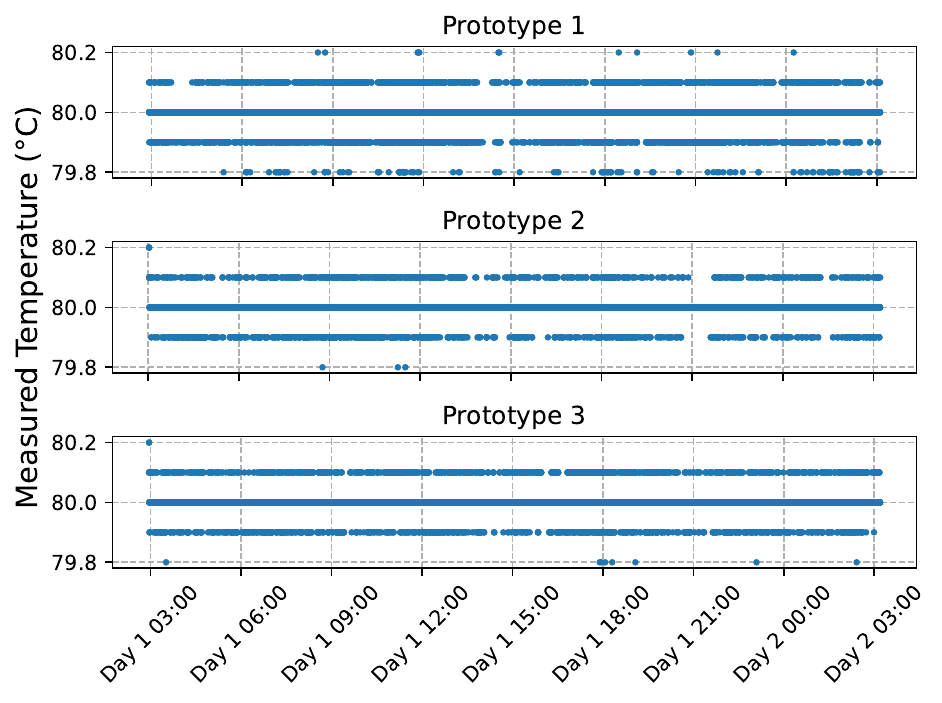}
    \caption{{Observed temperature measurements when the target temperature is set to \SI{80.0}{\celsius} in three different prototypes.}}
    \label{fig:temperature-stability}
\end{figure}

\subsubsection{Hardware IP Cores}
\label{sec:hardware-ip-cores}
We make use of several Xilinx IP components across our prototypes\omcr{.}

\noindent
\textbf{Xilinx XDMA PCIe Engine.} DRAM experiments can result in terabytes of data that is analyzed by the user program that runs on the host machine. Therefore, it is important to provide a high-speed host machine $\leftrightarrow$ FPGA board interface. We implement PCIe as the high-speed interface in our prototypes as it has high throughput (several Gb/s~\cite{pciespec}) and it is widely available among popular FPGA boards. We use the Xilinx DMA/Bridge Subsystem for PCIe Express IP~\cite{xdma} to communicate with the host machine over the PCIe interface. The IP is responsible for bringing up the PCIe connection with the attached host machine, and driving the physical PCIe signals with the data it receives from \X{} hardware {(data read from DRAM using \texttt{READ} commands)} over the standard AXI4-Stream interface~\cite{axispec}.

\noindent
\textbf{Xilinx DDR3/4 PHY.} We issue DRAM commands to the Xilinx DDR3/4 PHY IP~\cite{ultrascaleddr} over a low-level DFI interface. The IP receives multiple DRAM commands per FPGA clock cycle and performs the required operations to drive the physical DRAM interface signals (e.g., DQ, DQS) at high speeds, such as serializing multiple DRAM commands it receives to a higher frequency domain (e.g., {the} DDRx interface). By default, the IP {imposes} a constraint on the ordering of DDRx \texttt{READ} and \texttt{WRITE} commands. We modify the IP's Verilog source to remove this constraint.

FPGA designs typically use a DFI-like interface with a manufacturer-provided IP to communicate with memory devices~\cite{intelphy,xilphy}. DRAM Bender can be easily adapted to support other memory standards by interfacing with manufacturer-provided IPs built for these memory standards {(e.g., HBM2~\cite{xilhbm}, RLDRAM~\cite{intelrldram})}. {In fact, a recent work~\cite{olgun2023hbm} adapts \X{} to interface with modern HBM2 chips.}

\Copy{R1/7a}{\tcadreva{The IPs we use in our prototypes implement standardized interfaces to communicate with {the} \X{} hardware.\changev{\ref{q:r1q7}} \X{} can be relatively easily ported to new prototypes that use different IP {(e.g., open source IP)}, given that they use the same common standard interfaces, e.g., AXI4, PHY interface. In particular, Intel provides a PCIe IP~\cite{intelpcie} and an external memory interface IP~\cite{intelphy} for developers to use in their Intel FPGA designs. These IPs expose hardware interfaces similar to Xilinx's PCIe and DDR3/4 PHY IPs.}}

\subsubsection{DDR3 Changes}
\label{sec:ddr3-modifications}

\atb{To show that \X{} can be easily extended to support other DRAM interfaces and FPGA boards, we summarize the changes required {for \X{} to support DDR3} using the Xilinx ZC706 board~\cite{zc706}. First, we modify the \emph{DRAM interface adapter} (Section~\ref{sec:dram-interface-adapter}) to correctly translate from DRAM pipeline signals to Xilinx DDR3 PHY~\cite{virtex7mig} signals. Second, The PCIe connection is 0.6$\times{}$ slower in the ZC706 board (5.0 GT/s) compared to the other boards we use for DDR4 (8.0 GT/s). As a consequence, the AXI interface between the frontend and the Xilinx XDMA IP has a twice as narrow (128-bit) data bus. We make small modifications to the \emph{frontend} module (Section~\ref{sec:memory-elements}) to account for this change in the data bus. Third, we slightly modify the \texttt{sendData} function in the API to correctly partition instructions into packets sent over the PCIe bus. Overall, our changes introduce only \textbf{230} additional lines of Verilog code in the hardware description and \textbf{30} additional lines of C++ code in the API.} {Our Github repository includes all ZC706 DDR3 design sources~\cite{self.github}.}

\subsection{\lois{Experiment Workflow}}
\label{sec:experiment-workflow}

\Copy{R1/5b}{
\atb{To show how instructions and data flow between \X{} components during an experiment, w}\lois{e describe a typical experiment workflow when using the \X{} infrastructure. \changev{\ref{q:r1q5}}\tcadreva{Figure~\ref{fig:workflow-chart} summarizes the workflow.}}}

\begin{figure}[h]
    \Copy{R1/5bfig}{
    \centering
    \includegraphics[width=0.8\linewidth]{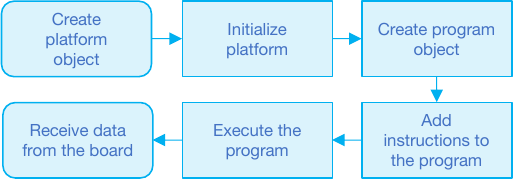}
    \caption{\tcadreva{\X{} experiment workflow}}
    \label{fig:workflow-chart}
    }
\end{figure}

\Copy{R1/5c}{
\tcadreva{The user 1) creates a platform object, 2) initializes the \X{} platform using the \texttt{platform.initialize()} function, 3) creates a program object, 4) adds instructions to the program object using \texttt{append<Instruction>} functions, 5) programs \X{} by supplying the program object to the \texttt{platform.execute()} function, and 6) retrieves data from the FPGA board by using the \texttt{platform.receiveData()} function to analyze output data.}}

\section{\tcadreva{Use Cases}}
\label{sec:case-studies}
\lois{To demonstrate that \X{} is a versatile DRAM testing infrastructure,} we conduct {three} {new} case studies {that result in new observations}: \One we investigate the effects of {the order in which the two aggressor rows are activated and precharged ({i.e.,} the interleaving pattern of activations to the aggressor rows)} on \lois{RowHammer}~\cite{kim2014flipping,mutlu2023fundamentally}{, \Two we investigate the effects of data patterns on RowHammer and show that \X{} can be used to uncover {new} insight\lois{s} {into the RowHammer vulnerability}}\lois{, and} \Three we {demonstrate for the first time that} processing-{using}-memory {capability exists} in off-the-shelf {DDR4 chips}. {The source code for all three case studies is available on our Github repository~\cite{self.github}.}

\noindent
\textbf{\lois{Experimental Setup}.} In all case studies, we use Alveo U200~\cite{Alveo} boards that are \om{set up} as described in \lois{Section}~\ref{sec:prototypes}. We keep the \lois{temperature of the} DRAM chips that we test at \SI{50}{\celsius}. We disable all DRAM self-regulation events (e.g., DRAM Refresh \cite{jedec2017ddr4}) to \One minimize all possible sources of interference \cite{kim2020revisiting}, and \Two disable proprietary in-DRAM RowHammer protection mechanisms (e.g., Target Row Refresh~\cite{frigo2020trrespass,hassan2021utrr}). We disable all forms of rank-level error-correction codes (ECC) to make sure ECC does not obscure RowHammer-induced bit-flips. Table~\ref{table:dram-devices} lists the DRAM {modules} from three different manufacturers that we use in our experiments. \atb{We test one memory module (8 DRAM chips) from each manufacturer.}

\begin{table*}[!t]
\centering
\caption{{DRAM {modules} tested in {all three} case studies}}
\label{table:dram-devices}
\begin{tabular}{|c|c|c|c|c|c|c|}
\hline
           \textbf{Manufacturer} & \textbf{{Module Part Number}} & \textbf{{Chip Part Number}} & \begin{tabular}[c]{@{}l@{}}\textbf{Standard}\end{tabular} & \begin{tabular}[c]{@{}l@{}}\textbf{\atb{Chip Density}}\end{tabular} & \begin{tabular}[c]{@{}l@{}}\textbf{Org.}\end{tabular} \\ 
           \hline \hline
\textbf{{Micron} (A)} & {MTA18ASF2G72PZ-2G3B1QG} & {MT40A2G4WE-083E:B} & DDR4 RDIMM & 8Gb & 1R$\times$4 \\ \hline
\textbf{{Hynix} (B)} & {KVR24R17S8/4} & {H5AN4G8NAFR-UHC} & DDR4 RDIMM & 4Gb & 1R$\times$8 \\ \hline
\textbf{{Samsung} (C)} & {F4-2400C17S-8GNT} & {K4A4G085WF-BCTD} & DDR4 UDIMM & 4Gb & 2R$\times$8 \\ \hline
\end{tabular}
\end{table*}

\subsection{Study \#1: RowHammer\omcr{:} {Interleaving Pattern of Activations}}
\label{sec:case-study-1}

{To demonstrate \X{}'s ease of use for developing new experiments that uncover new insights, we conduct a comprehensive experiment to study the effects of the interleaving pattern of activations to the aggressor rows on RowHammer bit-flips.}
{\changev{\ref{q:r1q5}}\tcadreva{Figure~\ref{fig:rowhammer-description} {depicts} the double-sided RowHammer attack~\cite{kim2014flipping,kim2020revisiting,orosa2021deeper,seaborn2015exploiting} in our case study.} We pick five consecutive rows V1, A1, V2, A2, V3, and rapidly activate the aggressor rows (A1 and A2) and record the bit-flips {observed in} the victim rows (V1, V2 and V3). {The interleaving pattern describes how many times one aggressor row is hammered before switching to the hammering of the other aggressor row in one iteration of the attack.}} 

\begin{figure}[h]
    \Copy{R1/5afig}{
    \centering
    \includegraphics[width=0.49\textwidth]{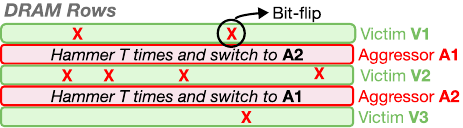}
    \caption{\tcadreva{Double-sided RowHammer {attack}}}
    \label{fig:rowhammer-description}
    }
\end{figure}

\noindent
\textbf{Methodology.} We use the checkered data pattern as it is reported to have the highest average RowHammer error coverage across DDR4 modules from three manufacturers in~\cite{kim2020revisiting}. This data pattern initializes {aggressor {rows} A1 and A2 {(see Figure~\ref{fig:rowhammer-description})}} with ``0xAA'' values and {victim {rows} V1, V2, and V3} with ``0x55'' values.
We activate {aggressors} in an interleaved manner. {In each iteration}, we {1)~}activate {the first} aggressor row $T$ times, {and 2)~activate the second aggressor row $T$ times}. We sweep the value of $T$ and observe {the number of bit-flips {in} the {victim rows}}. {We issue a total number of 1M ACT commands regardless of the value of $T$ (e.g., the attack finishes in {8 iterations} if $T = 64K$ and in {512K iterations} if $T=1$).} We create a double-sided \lois{RowHammer} attack program using the \X{} API. Listing~\ref{list:rowhammer-code} depicts the key code {portion} that performs {{an iteration {of}}} the attack{, which we repeatedly perform until we issue 1M ACT commands}. We sweep $T$ through the range [1, 64K], incrementing by a power of two {after each iteration} (i.e., we set $T$ to 1, 2, 4, ..., 64K) to perform the double-sided \lois{RowHammer} attack in a sequential ($T = 1$), and in a cascaded manner ($T > 1$). 

\vspace{3mm}
\begin{minipage}[c]{0.95\linewidth}
\begin{lstlisting}[caption={{\X{} program that performs {an iteration} of the double-sided RowHammer attack}},captionpos=b,label={list:rowhammer-code}]
p.appendLI(hammerCount, 0);
p.appendLabel("HAMMER1");
p.appendACT(bank, false, |\textcolor{red}{A1}|, false, |\textcolor{magenta}{tRAS}|);
p.appendPRE(bank, false, false, |\textcolor{magenta}{tRP}|);
p.appendADDI(hammerCount, hammerCount, 1);
p.appendBL(hammerCount, |\textcolor{red}{T}|, "HAMMER1");
p.appendLI(hammerCount, 0);
p.appendLabel("HAMMER2");
p.appendACT(bank, false, |\textcolor{red}{A2}|, false, |\textcolor{magenta}{tRAS}|);
p.appendPRE(bank, false, false, |\textcolor{magenta}{tRP}|);
p.appendADDI(hammerCount, hammerCount, 1);
p.appendBL(hammerCount, |\textcolor{red}{T}|, "HAMMER2");
\end{lstlisting}
\end{minipage}
\vspace{3mm}

We use two metrics to demonstrate \lois{the effects of {the interleaving pattern of activations} in a RowHammer attack}: \One \emph{average bit-flips per row} is the number of bit-flips in a DRAM bank averaged across all DRAM rows, \Two \emph{$HC_{first}$} is the number of \texttt{ACT} commands issued \lois{per aggressor row} in a \lois{RowHammer} attack to induce the first bit-flip in a particular DRAM row.

\noindent
\textbf{Results.} {Figures~\ref{fig:bit-flips-normalized-abs} and~\ref{fig:bit-flips-normalized}} \lois{show} the number of bit-flips we observe in different modules as we sweep $T$ from 1 to 64K.\footnote{To {crisply demonstrate the trends}, we leave out some of the T values from the plot. The trend we observe across those follow the trend {in} the figure.} Each bar represents the number of bit-flips we observe {in} one victim row (e.g., gold bar is the sandwiched row, V2) normalized to the number of bit-flips we observe {in} the same row at \lois{$T = 64K$}, for all rows in a DRAM bank. 

\begin{figure}[h]
    \centering
      \includegraphics[width=0.49\textwidth]{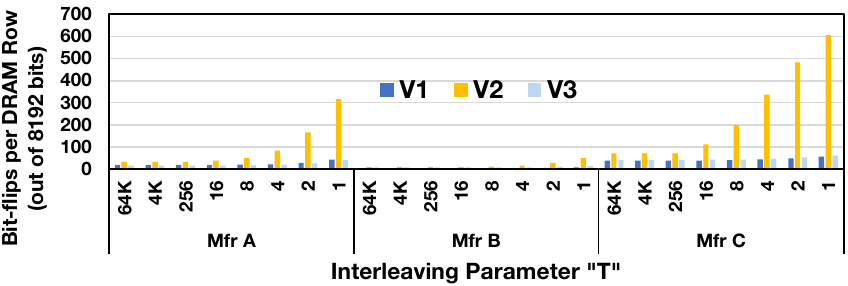}
    \caption{{{Effect of the interleaving pattern {on bit-flip rate}, {as} dictated by parameter ``$T$''}}}
    \label{fig:bit-flips-normalized-abs}
\end{figure}

\begin{figure}[h]
    \centering
      \includegraphics[width=0.49\textwidth]{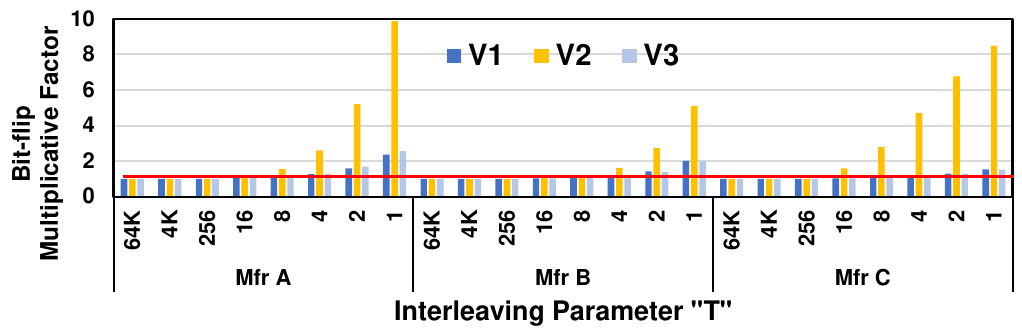}
    \caption{{Effect of the interleaving pattern {on bit-flip rate}, {as} dictated by parameter ``$T$'' (normalized to $T=64K$)}}
    \label{fig:bit-flips-normalized}
\end{figure}

We make two {new} observations from Figure~\ref{fig:bit-flips-normalized}. {First,} the {number} of bit-flips {in} the sandwiched row \lois{(V2)} significantly {increases} as $T$ \lois{approaches} one. Since the total {number} of \texttt{ACT} commands we issue (i.e., hammer count) remains the same, this \lois{observation} suggests that the {interleaving pattern} of the \texttt{ACT} commands {to the two aggressor rows} in a double-sided \lois{RowHammer} attack greatly affects the \atb{effectiveness} of the attack. We observe on average across all double-sided victim (V2) rows {(31.9, 9.9, 71.2) bit-flips at $T = 64K$}, and {(314.8, 50.7, 604.9) bit-flips at $T = 1$} for manufacturers (A, B, C) respectively. {The double-sided RowHammer attack more strongly resembles a single-sided attack as $T$ increases. Prior works (e.g.,~\cite{seaborn2015exploiting,aweke2016anvil}) show that single-sided attacks are less effective in inducing RowHammer bit-flips than double-sided attacks, and this could explain why our attack's effectiveness degrades as $T$ increases.} 

{Second,} \lois{we} observe that the {number} of bit-flips induced {in} the nearby victim rows, V1 and V3, also increase{s} as $T$ \lois{approaches} one. We hypothesize that a combination of two phenomena {explains} {{this} second observation}: \One V1 and V3 become less susceptible to \lois{RowHammer-induced} bit-flips as their immediate neighbor aggressor row is hammered in a cascaded manner (i.e., $T$ \lois{approaches} 64K). A cascaded {activation} pattern {causes} the aggressor row to stay precharged for a longer time (one aggressor stays precharged while the other aggressor is being hammered). This might reduce the overall electron injection {rate} between the aggressor and victim rows~\cite{walker2021ondramrowhammer,gautam2019row} {and thus} the \lois{RowHammer} vulnerability of the victim row. \Two {T}he \lois{RowHammer} blast radius effect~\cite{kim2014flipping}, in which an aggressor row can induce bit-flips \om{in} victim rows that are not its immediate neighbors (e.g., A2 and V1), is strengthened as the attack alternates between aggressor rows \lois{more frequently} (i.e., $T$ \lois{approaches} one).

{Figures~\ref{fig:rowhammer-hc-first-abs} and~\ref{fig:rowhammer-hc-first}} \lois{show} the {number} of \texttt{ACT} commands \lois{that is required} to observe the first \lois{RowHammer} bit-flip \lois{($HC_{first}$)} in the sandwiched DRAM row (V2) \lois{for} each DRAM \lois{manufacturer} \lois{using} different \lois{values of} $T$. For each $T$, we find the row with the smallest $HC_{first}$ in a DRAM bank. We observe {an $HC_{first}$ of} {99K, 80K, 16K at $T = 1$, and 130K, 108K, 23K at $T = 64K$} for manufacturers A, B, C, respectively. {Hence, the interleaving} pattern of the double-sided \lois{RowHammer} attack \lois{affects} the $HC_{first}$ of a DRAM row. As the attack's {interleaving} pattern becomes more cascaded (i.e., $T$ \lois{approaches} {64K}), $HC_{first}$ increases. \lois{RowHammer} defense mechanisms {(e.g.,~{\mitigatingRowHammerAllCitations{}})} may leverage this observation to render double-sided \lois{RowHammer} attacks less effective by scheduling the aggressor row accesses in a cascaded manner to the DRAM module.

\begin{figure}[h]
    \centering
    \includegraphics[width=0.49\textwidth]{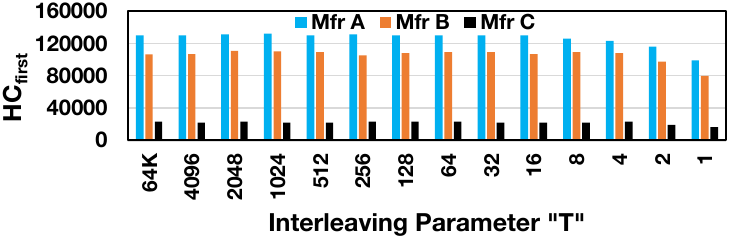}
    \caption{{{Effect of the interleaving pattern {on $HC_{first}$, as} {dictated by parameter ``$T$''}}}}
    \label{fig:rowhammer-hc-first-abs}
\end{figure}

\begin{figure}[h]
    \centering
    \includegraphics[width=0.49\textwidth]{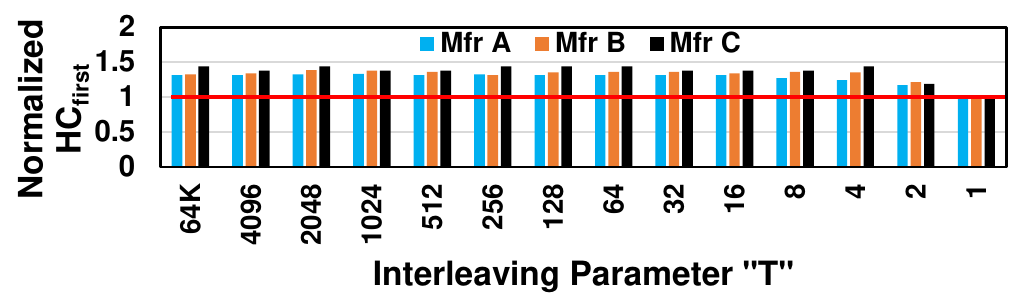}
    \caption{{Effect of the interleaving pattern {on $HC_{first}$, as} {dictated by parameter ``$T$''} (normalized to $T = 1$)}}
    \label{fig:rowhammer-hc-first}
\end{figure}

\noindent
\textbf{\atb{Conclusion.}} {Based on our new observations enabled by \X{},} \lois{we conclude that the {interleaving} pattern {of} the aggressor row {activations} in a double-si{d}ed RowHammer attack significantly affects 1) the {number} of RowHammer-induced bit-flips, and 2) the number of ACT commands required to induce the first RowHammer-induced bit-flip in a DRAM bank.}

\subsection{{Study \#2: RowHammer\omcr{:} Data Patterns}}
\label{sec:case-study-2}
{To {demonstrate that} \X{}'s nonrestrictive data interface is useful {for} generating new insights, we conduct a {comprehensive} experiment to study the {effect} of data patterns on RowHammer-induced bit-flips.}

\noindent
\textbf{Methodology.} We {run a \X{} test where we perform} {a} double-sided RowHammer attack on a victim row and observe the bit-flips that occur in a DRAM cache block (512 bits) in the victim row. {Our test code \One initializes the aggressor rows using either 1) a \emph{SoftMC} or 2) a \emph{\X{} data pattern}, \Two conducts double-sided RowHammer on the victim row, \Three records the positions of the bits that flip in one cache block in the victim row.} {We run two variations of the test code to observe the bit-flips induced by 1) SoftMC data patterns and 2) \X{} data patterns. First, we use {the limited} 256 8-bit data patterns supported by SoftMC, used in many prior works to conduct Rowhammer experiments~\cite{kim2020revisiting,kim2014flipping,hassan2021utrr,luo2023rowpress,orosa2022spyhammer,orosa2021deeper}}. {Second, we use the 256 randomly{-}generated 512-bit {\X{}} data patterns.} {We run the test code 100 times} and aggregate all bit-flips across the 100 runs of the experiment to collect the complete set of RowHammer failures (e.g., a bit-flip that occurs at least once during the 100 rounds of testing is recorded as a bit-flip). We test 24 randomly selected DRAM cache blocks in 24 DRAM rows from each DRAM module to maintain a reasonable testing time. We repeat {each test code variation} twice, once by initializing the victim row with an all-zeros and once by initializing the victim row with an all-ones data pattern, to cover cells that flip from $1-0$ and $0-1$.

\noindent
{\textbf{Results.} We identify at least one additional cell that flips from either $1-0$ or $0-1$ when we initialize the aggressor rows with random 512-bit data patterns compared to when we initialize the aggressor rows with SoftMC data patterns on every victim row that we test in modules from all three DRAM manufacturers. We conclude that \X{} can identify RowHammer-susceptible DRAM cells that cannot be identified using {the} SoftMC data patterns used in prior work. This information could be useful for strengthening RowHammer attacks that need to induce bit-flips {in} a particular cell~\cite{kwong2020rambleed}.}

\subsection{Study \#{3}: In-DRAM Bitwise Operations}
\label{sec:case-study-3}
Recent works propose processing-{using}-memory (PIM) techniques that can greatly improve system performance and reduce energy consumption~\pimcitations{}. \lois{A recent work demonstrates that bitwise AND/OR operations} can be performed in off-the-shelf DDR3 chips~\cite{gao2019computedram}. \lois{However,} it is unclear if {commonly-used} newer{-}generation off-the-shelf DDR4 chips support these {operations}. \lois{Our goal in this section is to {leverage \X{}'s support for fine-grained control over DRAM timing parameters and} {demonstrate that} bitwise AND/OR operations can also be performed in off-the-shelf DDR4 chips.}
\lois{In our experiments, we perform bitwise operations by issuing a sequence of \texttt{ACT}$\rightarrow{}$\texttt{PRE}$\rightarrow{}$\texttt{ACT} commands with violated timing parameters. We characterize our DRAM {modules} to obtain the {bit error rate (i.e., the {fraction} of bitlines that \emph{cannot} correctly perform the bitwise operation)} of the performed bitwise operations.}

\noindent
\textbf{Majority Operation.} To perform bitwise AND/OR operations, we {use} the \emph{majority} function across DRAM cells on the same bitline~\cite{seshadri2017ambit} in a small portion of a DRAM array. {We refer the reader to ComputeDRAM~\cite{gao2019computedram} for a detailed description {of} how real chips can perform majority operations.}

\noindent
\textbf{Methodology.} We perform bitwise AND/OR operations by issuing DRAM command sequences to DRAM modules {in quick succession by violating standard DRAM timing parameters}. Listing~\ref{list:majority-code} depicts the key code block that we use to generate \X{} programs for our experiment.

\vspace{3mm}
\begin{lstlisting}[caption={{\X{} {code segment} to perform {a bitwise} majority operation}},captionpos=b,label={list:majority-code}]
p.appendACT(BANK, false, R1, false, |\textcolor{red}{N}|);
p.appendPRE(BANK, false, false, |\textcolor{red}{M}|);
p.appendACT(BANK, false, R2, false);
\end{lstlisting}
\vspace{3mm}

We send the first \texttt{ACT} command to row {address} {R1}, and the second \texttt{ACT} command to row {address} {\emph{R2}}. {We select R1 and R2 such that they correspond to the second and the third row in a contiguous set of four DRAM rows (e.g., row addresses 0, 1, 2, 3), which we call a \emph{DRAM segment}. For example, the first row in the second DRAM segment in a bank has the row address $4$. Thus, R1 and R2 are $5$ and $6$ when we want to perform bitwise AND/OR in the second segment.} {In a segment,} we initialize {the \emph{second row}} with \emph{all-ones} and {place the operands in the first and the third rows to perform an OR operation}, and initialize {the \emph{first row}} with \emph{all-zeros} and {place the operands in the second and the third rows to perform an AND operation~\cite{gao2019computedram}}. {We test a total number of 8K segments in one DRAM bank and calculate the bit error rate (BER) of the bitwise AND/OR operations for each segment.} We sweep the values of $tRAS$ {(N in Listing~\ref{list:majority-code})} and $tRP$ {(M in Listing~\ref{list:majority-code})} from \SI{1.5}{ns} to \SI{15}{ns} with increments of \SI{1.5}{ns}, and test a total {number} of 100 different $tRAS$ and $tRP$ combinations.

\noindent
\textbf{Results.} {We make the key observation from our experiments that we} can perform valid majority AND/OR operations {(i.e., we observe that multiple rows are activated with the execution of the \texttt{ACT}$\rightarrow{}$\texttt{PRE}$\rightarrow{}$\texttt{ACT} command sequence)} on DRAM chips from one DRAM \yct{manufacturer} {(i.e., SK Hynix)} when we use the following ($tRAS$, $tRP$) timing combinations: (\SI{1.5}{ns}, \SI{1.5}{ns}), (\SI{1.5}{ns}, \SI{3.0}{ns}), and (\SI{3.0}{ns}, \SI{1.5}{ns}).

Figure~\ref{fig:heterogeneous-and-or-success} depicts the proportion of rows that can perform AND/OR operations with smaller BER than what is displayed on the x-axis. Each bar is composed of three stacks: \One the blue stack represents the {number} of segments that can perform AND operations, but not OR operations, \Two the green stack represents the {number} that can perform OR operations, but not AND operations, and \Three the gold stack represents the {number} that can perform both AND/OR operations{, all proportional to the number of segments that can correctly perform either one or both of the operations}. There is heterogeneity in the {BER} in terms of the operation (AND/OR) being performed. We find that \One AND operations produce more reliable results {than} OR operations. {35 segments} support \emph{only} AND operations at less than 3\% BER and the smallest observed BER is 1.9\% for AND operations. \Two The number of segments that support AND/OR operations {significantly increases} as BER increases: 160 segments support AND/OR with $<5$\% BER, whereas 4546 segments support AND/OR with $<10$\% BER.

\begin{figure}[h]
    \centering
    \includegraphics[width=0.49\textwidth]{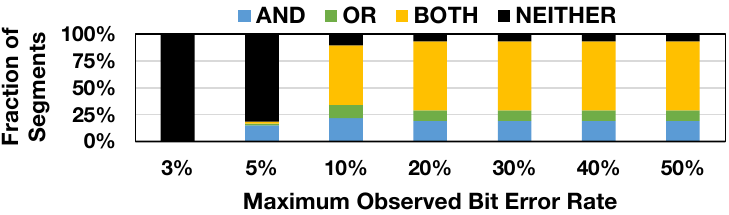}
    \caption{{The {fraction} of DRAM segments that support AND, OR, both (AND and OR) {of}{, and neither} of the operations.}}
    \label{fig:heterogeneous-and-or-success}
\end{figure}

\noindent
\textbf{Conclusion.} We conclude that new-generation DDR4 {chips} support in-DRAM bitwise AND/OR operations. Although we do not find any DRAM segments that support AND/OR operations with 0\% BER, our results {align} with prior work {on DDR3 chips}~\cite{gao2019computedram}. {We make a new observation {that t}here is} heterogeneity in BER in terms of the bitwise operation performed, which can be leveraged by approximate computing techniques to perform more accurate computation~\cite{koppula2019eden}. For example, a workload could place its bitwise AND operations' operands {in}to segments {with} a smaller BER for AND operations.

\subsection{\tcadreva{Research Enabled by \X{}}}
\label{sec:research-enabled}
\changev{\ref{q:r1q9}}
\om{\X{} has already enabled {several} prior work{s} on {read disturbance} characterization~\cite{orosa2021deeper,kim2020revisiting,yaglikci2022understanding,luo2023rowpress,orosa2022spyhammer,hassan2021utrr,frigo2020trrespass}{,} evaluation of existing RowHammer mitigations~\cite{frigo2020trrespass,hassan2021utrr}, true random number generation~\cite{olgun2021quactrng}, {uncovering undocumented functionality in DRAM~\cite{yaglikci2022hira},} and {approximate DRAM for efficient} deep neural network {inference}~\cite{koppula2019eden}. Parts of \X{}'s hardware design {were} reused in developing an end-to-end {processing-in-DRAM} framework~\cite{olgun2021pidram} {and an HBM2 DRAM testing infrastructure~\cite{olgun2023hbm}}.}

\om{All of these works have provided significant contributions and advanced the state-of-the-art in their respective areas. Particularly, the works that evaluate RowHammer mitigation mechanisms used in real DDR4 DRAM chips demonstrated that these DRAM chips are \emph{not} RowHammer-free, contrary to major DRAM manufacturers' claims~\cite{frigo2020trrespass,hassan2021utrr} {and thus enabled significant changes in industry~\cite{mutlu2023fundamentally}}.}

\new{The artifacts of two of the prior works that use \X{}, U-TRR~\cite{hassan2021utrr} and QUAC-TRNG~\cite{olgun2021quactrng}, are already freely and openly available~\cite{utrr.github,quactrng.github}. The open source artifacts of other prior work~\cite{frigo2020trrespass,orosa2021deeper,kim2020revisiting,yaglikci2022understanding,koppula2019eden,yaglikci2022hira} that used \X{} are work-in-progress and we plan to release them as part of our future work.}

\om{We believe that these prior works demonstrate the versatility and ease of use of \X{} in enabling impactful research. {W}e hope that {such versatility and ease of use} accelerates the adoption of \X{} by researchers in academia and industry to perform experiments on state-of-the-art DRAM chips.} 

\section{{New Research Directions}}
\label{sec:new-research-directions}

\X{} users can conduct studies on state-of-the-art DRAM chips with no interface restrictions. We discuss two concrete example research directions.

\subsection{{Characterization of RowHammer}}
{\X{} enables open study of DRAM security and reliability issues using state-of-the-art chips. \X{} openly and freely enables studies that evaluate and verify the security guarantees provided by RowHammer mitigation mechanisms implemented in real DDR4 devices (similar to~\cite{frigo2020trrespass, hassan2021utrr}). Since DDR4 DRAM is widely used in current and near-future computing systems, many systems can be vulnerable to attacks that can escalate user privileges, leak confidential data and perform denial of service if bit-flips are exploitable. It is important to rigorously understand the RowHammer problem~\cite{mutlu2023fundamentally} and other potential vulnerabilities that may lead to bit-flips (e.g., data retention errors) and openly evaluate any existing mitigations in order to develop effective and comprehensive solutions to RowHammer and other potential vulnerabilities. \X{} can enable new studies similar to~\cite{frigo2020trrespass, hassan2021utrr} as well as other studies that might discover other issues in state-of-the-art DRAM chips.}

\subsection{Power Consumption Studies}
\label{sec:power-consumption-studies}
Prior works~\cite{vogelsang2010understanding, ghose2018vampire} show that the existing DRAM power models are not accurate \om{for} two \om{key} reasons. First, manufacturer-defined guardbands, provided in data sheets, prevent one from understanding the real power consumption characteristics of DRAM chips and can be misleading regarding the power consumption variation across different manufacturers, die revisions, and technology nodes. Second, standard power consumption values, referred to as $IDD(N)$ and $ICC(N)$ values, do not cover the worst-case (i.e., peak) power consumption of real chips due to static and limited data patterns used during these measurements, as specified in technical documents~\cite{jedec-lpddr2, jedec-lpddr3, jedec2015lpddr4, jedec2017ddr4, jedec2015lpddr4, jedec2020lpddr5, jedec2020ddr5, micron_lpddr3, micron2007power, micron2017power}. Therefore, it is important for future research to understand a DRAM chip's power consumption under different conditions~\cite{ghose2018vampire, drampower}. 

Unfortunately, publicly available DRAM testing infrastructures are not capable of accurately measuring a DRAM's energy consumption. The state-of-the-art, SoftMC~\cite{hassan2017softmc}, has a limited capacity (8K) DRAM command queue {with} no control flow support for branch and jump {operations}, causing a given command sequence to be completed within less than \SI{10}{\micro\second}, after which a new command sequence must be transmitted from a host machine to the FPGA-based platform which can incur latency in the order of \SI{}{\milli\second}, depending on the host machine's state. Therefore, accurately measuring a DRAM chip's power consumption with SoftMC requires employing power measurement equipment with a \SI{}{\mega\hertz} sampling rate. However, even high-end current measuring equipment~\cite{keysighta} {have sampling rates of fewer than \SI{10}{\kilo\hertz}}~\cite{ghose2018vampire}.
To overcome this limitation, Ghose et al.~\cite{ghose2018vampire} report that they heavily modify SoftMC~\cite{hassan2017softmc} to allow the command sequence to be executed in an infinite loop. Unfortunately, their modifications are not publicly available and it is not clear if their test setup can be used for standards newer than DDR3.

\X{}'s programmer interface allows a user to create loops to repeat executing a command sequence as many times as needed for power measurement equipment\del{\footnote{{\del{We currently lack this equipment to perform the described power consumption study. We plan to incorporate the equipment as a part of \X{} in future work.}}}} to accurately capture the power consumption. We believe that \X{} will enable future studies on DRAM power consumption, which is a first-order concern on a wide spectrum of systems, including servers, mobile devices, and battery-backed edge devices. 

\section{Related Work}
\label{sec:related-work}

{\X{} enables experimental studies on state-of-the-art DRAM chips by providing an open, easy-to-use, and extensible DRAM testing infrastructure. We {already extensively discuss the shortcomings of existing open source DRAM testing infrastructures in Section~\ref{sec:existing-infrastructures}}. We now compare \X{} against other infrastructures and platforms.} 

Besides SoftMC~\cite{hassan2017softmc} and LRT~\cite{litex.github}, there exist other infrastructures used for testing DRAM chips~\cite{maosong2020characterize, advantest,
nickel, teradyne, futureplus,huang2000fpga, hou2013fpga, keezer2015fpga,bojnordi2012pardis}.
{\Copy{R2/1}{Unfortunately, those infrastructures suffer \atb{from} at least one of the following three shortcomings: 1)~support for limited and fixed test patterns~\cite{advantest,
nickel, teradyne, futureplus}, 
\changev{\ref{q:r2q1}}\tcadrevb{2)~poor usability due to lack of flexibility and open source code~\cite{huang2000fpga, hou2013fpga, keezer2015fpga},} and 3)~lack of programmability~\cite{bojnordi2012pardis}.} DRAM chips include a mechanism called {Built-In Self-Test (}BIST) (e.g.,~\cite{you1985self, querbach2014reusable,
querbacharchitecture, bernardi2010programmable, yang2015hybrid,
itoh2013vlsi, 
dreibelbis1998asic, agrawal1994builtin, dreibelbis1998processor, hakmi2007programmable, steininger2000testing, mccluskey1985techniques, mccluskey1985structures, agrawal1993tutorialapplications, agrawal1993tutorialprinciples}) to conduct fixed test patterns. Unfortunately, BIST 1)~provides only a limited number of fixed tests {that are not programmable}, {2)~lacks flexibility and visibility to perform many experiments, and~3)}~is implemented only in a limited set of DRAM chips~{\cite{hassan2017softmc}}. 
In contrast, as we demonstrate, \X{} provides an \atb{extensible design and an easy-to-use API. \X{} fully exposes the underlying DRAM interface to the user} to allow precisely crafting a desired access pattern{. \X{} users can} conduct tests using either a supported FPGA board and DRAM standard {out of the box} or {a new FPGA board and a new DRAM standard with} {relatively} simple modifications.}

{Prior works develop other {open source} FPGA-based frameworks for 1) evaluating processing-in-memory ({PIM}) techniques on real DRAM chips on real systems~\cite{olgun2021pidram}, 2) facilitating the emulation of PIM architectures and prototyping new PIM techniques~\cite{mosanu2022pimulator}, and 3) evaluating HBM-based processing-near-memory (PNM) architectures~\cite{zhang2020MEG}. Although useful in conducting real system evaluation of {PIM} and PNM techniques using contemporary workloads (e.g., performance, energy), unlike \X{}, these frameworks cannot be used to study the performance, reliability, and security characteristics of real, state-of-the-art DRAM chips.}

\section{Future Work and Limitations}
\label{sec:future-work}

\noindent
\atb{\textbf{Support For New DRAM Interfaces.} We leave {the} integration of new DRAM interfaces, such as DDR5~\cite{jedec2020ddr5}, to improve the versatility of \X{} to future work. DDR5 implements new commands to improve the performance and reliability of DRAM. In particular, the \texttt{RFM} command triggers a refresh management operation within the DRAM device, performing target row refresh (TRR) operations to alleviate RowHammer-induced bit-flips. Like other TRR mechanisms implemented in {the} current generation of DDRX interfaces, it is unclear if \texttt{RFM} can solve the RowHammer problem~\cite{frigo2020trrespass}. We hope to enable studies on \texttt{RFM} going forward as we integrate support for the DDR5 interface.}

\Copy{R1/7b}{\noindent
\changev{\ref{q:r1q7}}\tcadreva{\textbf{Support for New FPGA Boards.} \X{} is already prototyped on two different Xilinx boards and three different Bittware FPGA boards (Section~\ref{sec:prototypes}). Increasing the number of FPGA board prototypes is critical for \X{} to be {more} widely adopted as a DRAM testing infrastructure. More \X{} prototypes would allow a greater set of researchers with a limited number of FPGA boards in their inventories to use \X{} out of the box.}}

\Copy{R1/6}
{\noindent
\changev{\ref{q:r1q6}}\tcadreva{\textbf{Power Measurement Setup.}}
\tcadreva{A DDRx power measurement setup for DRAM Bender is work-in-progress. We aim to release a manual describing how to build a power measurement setup as part of future work in our Github repository~\cite{self.github}.}
\tcadreva{The planned power measurement setup is based on a prior work~\cite{ghose2018vampire} and consists of three major components. First, a \emph{DIMM riser board} (e.g., JET-5612~\cite{jet5612}) is placed between the DDRx socket on the FPGA board and the DDRx module under test. The riser board exposes the DDRx DIMM's voltage rails for current probing via shunt resistors. Second, a \emph{high-precision current sensor} (e.g., Keysight 34134A~\cite{keysighta}) connects to the shunt resistors and measures the current on the DDRx voltage rails. Third, a \emph{high-precision multimeter} (e.g., Keysight 34461A~\cite{keysightb}) is coupled with the current sensor to report current measurements.}}

\noindent
\atb{\textbf{Packetized Interfaces.} \om{A subset of the} current 3D stacked DRAM devices (e.g., HMC~\cite{HMC2}) implement \emph{packetized} interfaces that use high-speed serial links to communicate with the host processing unit. In such devices, the controller that communicates with the DRAM dies resides in the logic layer of the device. Therefore, memory packets specify operations at a higher level (e.g., \texttt{load} and \texttt{store}) compared to synchronous interfaces (e.g., \texttt{ACT}, \texttt{PRE}, \texttt{READ}). In an FPGA-based platform that supports such an interface, this would prevent \X{} from fully exposing the DRAM interface to the user as it would need to communicate with the DRAM device using higher-level operations. Hence, the packetized interface is a limiting factor in integrating support for \new{a subset of} current 3D stacked DRAM devices \new{(e.g., HMC)} in \X{}.}

\Copy{R1/2}{\noindent
\changev{\ref{q:r1q2}}\tcadreva{\textbf{Graphical User Interface}. A graphical user interface (GUI) could provide \X{} with a more intuitive and user-friendly interface for conducting experiments. We already open sourced \X{} on Github~\cite{self.github}. We believe that the open source repository will allow for a collaborative effort from many other researchers and users of the infrastructure to develop new valuable features, such as a GUI.}}

\section{Conclusion}
\atb{We develop a new open source DRAM testing infrastructure that is \om{versatile,} extensible, and easy to use. {Unlike} existing open source infrastructures, \X{} provides both \One a nonrestrictive interface to DRAM devices and \Two an \hh{extensible design}. Its nonrestrictive API allows users to issue DRAM commands in \om{an} arbitrary order and at arbitrary times. It comprises a modular design that \om{enables} easily integrat\om{ing} new DRAM interfaces and port\om{ing} to different FPGA boards. We demonstrate \hh{\X{}'s} \One versatility and ease of use by conducting {three} {new} studies \hh{that \om{show how \X{} successfully {enables} new insights {into the characteristics of modern DDR4} DRAM chip{s}}, \om{and} \Two extensibility by porting it to five FPGA boards \hh{with} either DDR3 or DDR4 support.} \om{We hope that with its versatility, ease of use, and extensible design, \X{} will be the mainstream infrastructure used in experimental {DRAM} studies and will be useful in developing new mechanisms and methodologies that improve DRAM security, reliability, performance{, and efficiency}.}}

\section*{Acknowledgments}
{We thank the anonymous reviewers of TCAD, ISCA 2022, and HPCA 2022 for feedback.} {We thank {the} SAFARI Research Group members for {valuable} feedback and the stimulating intellectual environment they provide. We acknowledge the generous gift funding provided by our industrial partners ({especially} Google, Huawei, Intel, Microsoft, VMware). This work was in part supported by {a} {Google Security and Privacy Research Award and the Microsoft Swiss Joint Research Center}.}


\begin{spacing}{0.80}
\footnotesize
\bibliographystyle{IEEEtranS}
\bibliography{newrefs,references}
\end{spacing}
\balance
\end{document}